\documentclass[prl,superscriptaddress,amsmath,amssymb,twocolumn]{revtex4-2}
\usepackage{natbib}
\usepackage{graphicx}         

\usepackage{textcmds}
\usepackage{dcolumn}          
\usepackage{bm}               
\usepackage{upgreek}
\usepackage{booktabs} 				
\usepackage{tabularx}
\usepackage{array}
\usepackage[colorlinks=true,urlcolor=blue,breaklinks=true]{hyperref}
\usepackage{xcolor}
\usepackage[a4paper,left=2cm,right=2cm]{geometry}
\usepackage{orcidlink}

\begin{document}

\title{Topological insulator based axial superconducting quantum interferometer structures}

\author{Erik Zimmermann\,\orcidlink{0000-0002-1159-2027}}
\email{erik.zimmermann@rwth-aachen.de}
\affiliation{Peter Gr\"unberg Institut (PGI-9), Forschungszentrum J\"ulich, 52425 J\"ulich, Germany}
\affiliation{JARA-Fundamentals of Future Information Technology, J\"ulich-Aachen Research Alliance, Forschungszentrum J\"ulich and RWTH Aachen University, Germany}

\author{Abdur Rehman Jalil\,\orcidlink{0000-0003-1869-2466}}
\affiliation{Peter Gr\"unberg Institut (PGI-9), Forschungszentrum J\"ulich, 52425 J\"ulich, Germany}
\affiliation{JARA-Fundamentals of Future Information Technology, J\"ulich-Aachen Research Alliance, Forschungszentrum J\"ulich and RWTH Aachen University, Germany}

\author{Michael Schleenvoigt\,\orcidlink{0000-0001-7977-7848}}
\affiliation{Peter Gr\"unberg Institut (PGI-9), Forschungszentrum J\"ulich, 52425 J\"ulich, Germany}
\affiliation{JARA-Fundamentals of Future Information Technology, J\"ulich-Aachen Research Alliance, Forschungszentrum J\"ulich and RWTH Aachen University, Germany}

\author{Jan Karthein\,\orcidlink{0009-0009-6887-8016}}
\affiliation{Peter Gr\"unberg Institut (PGI-9), Forschungszentrum J\"ulich, 52425 J\"ulich, Germany}
\affiliation{JARA-Fundamentals of Future Information Technology, J\"ulich-Aachen Research Alliance, Forschungszentrum J\"ulich and RWTH Aachen University, Germany}

\author{Benedikt Frohn\,\orcidlink{0000-0002-1404-9230}}
\affiliation{Peter Gr\"unberg Institut (PGI-9), Forschungszentrum J\"ulich, 52425 J\"ulich, Germany}
\affiliation{JARA-Fundamentals of Future Information Technology, J\"ulich-Aachen Research Alliance, Forschungszentrum J\"ulich and RWTH Aachen University, Germany}

\author{Gerrit Behner\,\orcidlink{0000-0002-7218-3841}}
\affiliation{Peter Gr\"unberg Institut (PGI-9), Forschungszentrum J\"ulich, 52425 J\"ulich, Germany}
\affiliation{JARA-Fundamentals of Future Information Technology, J\"ulich-Aachen Research Alliance, Forschungszentrum J\"ulich and RWTH Aachen University, Germany}

\author{Florian Lentz\,\orcidlink{0000-0002-8716-6446}}
\affiliation{Helmholtz Nano Facility (HNF), Forschungszentrum J\"ulich, 52425 J\"ulich, Germany}

\author{Stefan Trellenkamp}
\affiliation{Helmholtz Nano Facility (HNF), Forschungszentrum J\"ulich, 52425 J\"ulich, Germany}

\author{Elmar Neumann\,\orcidlink{0000-0001-9668-8281}}
\affiliation{Helmholtz Nano Facility (HNF), Forschungszentrum J\"ulich, 52425 J\"ulich, Germany}

\author{Peter Sch\"uffelgen\,\orcidlink{0000-0001-7977-7848}}
\affiliation{Peter Gr\"unberg Institut (PGI-9), Forschungszentrum J\"ulich, 52425 J\"ulich, Germany}
\affiliation{JARA-Fundamentals of Future Information Technology, J\"ulich-Aachen Research Alliance, Forschungszentrum J\"ulich and RWTH Aachen University, Germany}

\author{Hans L\"uth}
\affiliation{Peter Gr\"unberg Institut (PGI-9), Forschungszentrum J\"ulich, 52425 J\"ulich, Germany}
\affiliation{JARA-Fundamentals of Future Information Technology, J\"ulich-Aachen Research Alliance, Forschungszentrum J\"ulich and RWTH Aachen University, Germany}

\author{Detlev Gr\"utzmacher\,\orcidlink{0000-0001-6290-9672}}
\affiliation{Peter Gr\"unberg Institut (PGI-9), Forschungszentrum J\"ulich, 52425 J\"ulich, Germany}
\affiliation{JARA-Fundamentals of Future Information Technology, J\"ulich-Aachen Research Alliance, Forschungszentrum J\"ulich and RWTH Aachen University, Germany}

\author{Thomas Sch\"apers\,\orcidlink{0000-0001-7861-5003}}
\email{th.schaepers@fz-juelich.de}
\affiliation{Peter Gr\"unberg Institut (PGI-9), Forschungszentrum J\"ulich, 52425 J\"ulich, Germany}
\affiliation{JARA-Fundamentals of Future Information Technology, J\"ulich-Aachen Research Alliance, Forschungszentrum J\"ulich and RWTH Aachen University, Germany}

\hyphenation{}
\date{\today}

\begin{abstract}
Nanoscale superconducting quantum interference devices (SQUIDs) are fabricated in-situ from a single Bi$_{0.26}$Sb$_{1.74}$Te$_{3}$ nanoribbon that is defined using selective-area growth and contacted with superconducting Nb electrodes via a shadow mask technique. We present $h/(2e)$ magnetic flux periodic interference in both, fully and non-fully proximitized nanoribbons. The pronounced oscillations are explained by interference effects of coherent transport through topological surface states surrounding the cross-section of the nanoribbon.  
\end{abstract}
\maketitle
\section{Introduction}
On the way to topological quantum computing topological insulators (TIs) are promising candidates that gain more and more attention~\cite{he2019topological,breunig2022opportunities}. TIs are defined by their nontrivial topological properties that are invariant under small distortions~\cite{hasan2010colloquium,zhao2014topological,Ando13}. When combining TIs with an s-wave superconductor they are predicted to host so-called Majorana zero modes that may provide a robust and error reduced platform for quantum operations~\cite{Moore10,Kitaev03,Nayak08,Hyart13,alicea2012new}. Josephson junctions or interferometer structures formed of a TI in combination with an s-wave superconductor are used to investigate the interplay between both materials~\cite{fu2008superconducting,veldhorst2012optimizing,lucignano2013topological}.

\begin{figure*}[btp]
\centering
\includegraphics[width=\textwidth]{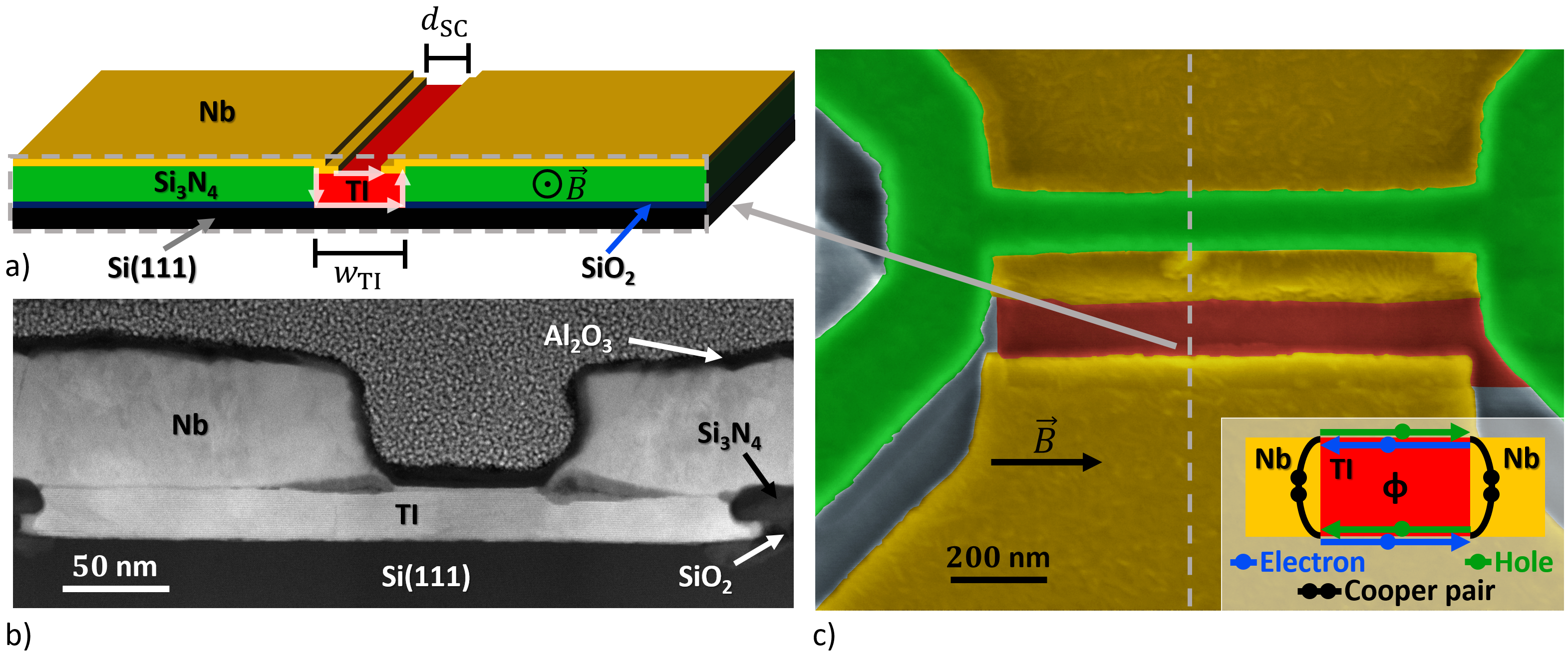}
\caption{a) Schematic illustration of the SQUID structure. The TI (red) is grown in a trench of SiO$_2$ (blue) and Si$_3$N$_4$ (green) on the Si(111) substrate (black). Then, Nb (yellow) is deposited on top with a shadow mask (not shown) leaving an uncovered area on top of the middle of the wire. The two interfering paths are marked as white arrows. The magnetic field is oriented along the nanoribbon. b) High angular annular dark field (HAADF) image acquired via scanning transmission electron microscope at the cross-section to inspect the TI epitaxial layer and the interfaces between Nb and TI. More details are provided in the Supplementary Material. c) False color scanning electron microscope (SEM) image showing a device from the top. The shadow mask is depicted in green. Since the Nb is deposited under an angle, the shadow is shifted downwards allowing a look on the junction without any need of tilting the sample. In the inset the concept of resonant Andreev reflections accross the cross section of the nanoribbon is sketched, schematically.}\label{Fig_TI_SQUID_Schema_SEM_TEM}
\end{figure*}

Weak-link Josephson junctions are made by bridging two closely spaced superconducting electrodes with a conductive material, which can be either metallic, semiconducting, or as in our case a topological insulator~\cite{likharev1979,schapers2001superconductor,schuffelgen2019selective,williams2012unconventional}. Figure~\ref{Fig_TI_SQUID_Schema_SEM_TEM}a) depicts a sketch of the device configuration investigated by us. A TI nanoribbon of width $w_\mathrm{TI}$ and thickness $d_\mathrm{TI}$ is contacted by in-situ deposited superconducting electrodes that approach the nanoribbon sideways from the top. Due to the lateral electrodes and the topologically protected surface states of the TI, this in theory results in a DC superconducting quantum interference device (SQUID) interferometer, as a short Josephson junction is formed on top and a longer one at the bottom including the side walls of the wire. The SQUID area is oriented perpendicular with respect to the area of the Josephson junctions. By applying a magnetic field along the nanoribbon axis it is expected that the resistance of the device is periodically modulated with a magnetic field period of $\Delta B=\phi_0/(w_\mathrm{TI} d_\mathrm{TI})$ with $\phi_0=h/2e$ the magnetic flux quantum. The device geometry is inspired by the GaAs/InAs core-shell nanowire based junctions presented by Haas \textit{et al.}~\cite{haas2018quantum}. The main difference is that in our device the gapped bulk and conducting surface behaviour originates in the TI properties, so that only a single material besides the superconducting electrodes is required.

In the following, the magnetotransport measurements of induced superconducting and non-superconducting interferometer structures are discussed. SQUID-like interference phenomena are presented which fit to the cross-sectional area of the ribbon that is enclosed by the surface states. Thus, the suggested device is a nano-scale SQUID that is suitable for Majorana zero mode detection in the future~\cite{veldhorst2012optimizing,lucignano2013topological}. Furthermore, it is shown that a coherent proximity effect has a striking effect on the oscillation amplitude, as induced superconducting samples show surface dominated transport.

\section{Experimental}
First, $10\,$nm SiO$_2$ and $25\,$nm Si$_3$N$_4$ are deposited by thermal oxidation and plasma-enhanced chemical vapor deposition, respectively, on a Si(111) wafer. The stack is patterned in a trench shape with a width of $200\,$nm - $300\,$nm by a resist process using electron-beam lithography and reactive-ion etching (RIE)~\cite{jalil2023selective}. Subsequently, a second stack of $300\,$nm SiO$_2$ and $100\,$nm Si$_3$N$_4$ is deposited. The topmost Si$_3$N$_4$ is then patterned as a shadow mask with a bridge width of $80\,$nm - $110\,$nm. Next, the SiO$_2$ is removed with hydrofluoric acid before growing the TI selectively in the trench via molecular-beam epitaxy~\cite{schmitt2022integration,kolzer2023supercurrent}. During the process the sample is rotated around the normal axis in order to ensure that the TI is grown homogeneously underneath the shadow mask. As TI the ternary composition of Bi$_{0.26}$Sb$_{1.74}$Te$_{3}$ is selected. In order to get the cleanest possible interface between TI and superconductor, an in-situ process for electrode fabrication is used. Nb is deposited from an angle without rotation of the sample, leaving a gap in the middle of the TI ribbon due to the Si$_3$N$_4$ shadow mask. The dimensions of the shadow mask ensure that the superconductor still covers the outer parts of the TI ribbon for electrical contacting. As a final step, a capping layer of $5\,$nm Al$_2$O$_3$ is deposited under rotation in order to prevent the device from oxidation when exposing it to ambient conditions. Finally, the Nb electrodes are shaped by SF$_6$ using RIE. In Fig.~\ref{Fig_TI_SQUID_Schema_SEM_TEM}b) a cross-sectional view of the sample taken by transmission electron microscopy (TEM) is shown. A top-view scanning electron micrograph is presented in Fig.~\ref{Fig_TI_SQUID_Schema_SEM_TEM}c). The fabrication process is depicted in detail in the Supplementary Material.

The magnetotransport measurements are performed in a $^3$He cryostat with a base temperature of $0.4\,$K. A superconducting magnet allows to apply an axial magnetic field. The electrical measurements are conducted in a quasi 4-terminal setup using AC lock-in techniques for the current biased measurements at zero magnetic field as well as standard DC resistance measurements for the magnetic field dependent measurements. In the Supplementary Material, the normal transport properties of the material are analysed using a reference Hall bar. Here, a total two-dimensional (2D) charge carrier concentration of $n_{\text{2D}}=3.72\cdot 10^{12}\,$cm$^{-2}$ and a mobility of $\mu=380\,$cm$^2$/Vs at $1.3\,$K are derived. The high charge carrier density as well as the low average mobility indicate that besides the topological surface states also bulk states contribute to the electrical transport~\cite{weyrich2016}.

Table~\ref{tab_Nb_SQUIDs} lists the relevant sample dimensions of four characterized samples. From the TEM measurements, a TI thickness $d_\mathrm{TI}$ of about $24\,$nm is determined. Among the identically fabricated devices, for sample A the phase-coherent proximity effect in the TI leads to a vanishing resistance due to the overlap of the superconducting wave functions. In contrast, samples B to D show a non-vanishing resistance. This is attributed to the microscopic structure of the TI, as sample specific effects like the distribution of grain boundaries may hinder an overlap of the induced superconducting wave functions in samples B to D, and hence, create a finite resistance.
\begin{table}[hbtp]\centering
\caption{Parameters of SQUID samples. The distance between the Nb electrodes $d_\text{sc}$ and the width of the TI nanoribbons $w_\text{TI}$ that are indicated in Fig.~\ref{Fig_TI_SQUID_Schema_SEM_TEM} a) are compared. The calculated periods $\Delta B_\text{cal}$ derived from the geometries of the SQUIDs are compared to the experimentally determined periods $\Delta B_\text{exp}$.}
\begin{tabular}{c|cc|cc}

Sample & $d_\text{sc}\,$(nm) & $w_\text{TI}\,$(nm) & $\Delta B_\text{cal}\,$(mT)& $\Delta B_\text{exp}\,$(mT)\\ 
\hline
A & 80 & 300 & 285 & 205\\ 
\hline
B & 110 & 300 & 285 & 345\\ 

C & 110 & 200 & 427 & 466\\ 

D & 80 & 200 & 427 & 488\\ 
\end{tabular} 
\label{tab_Nb_SQUIDs}
\end{table}

\section{Superconducting Interferometer}
\begin{figure}[btp]
\centering
\includegraphics[width=0.47\textwidth]{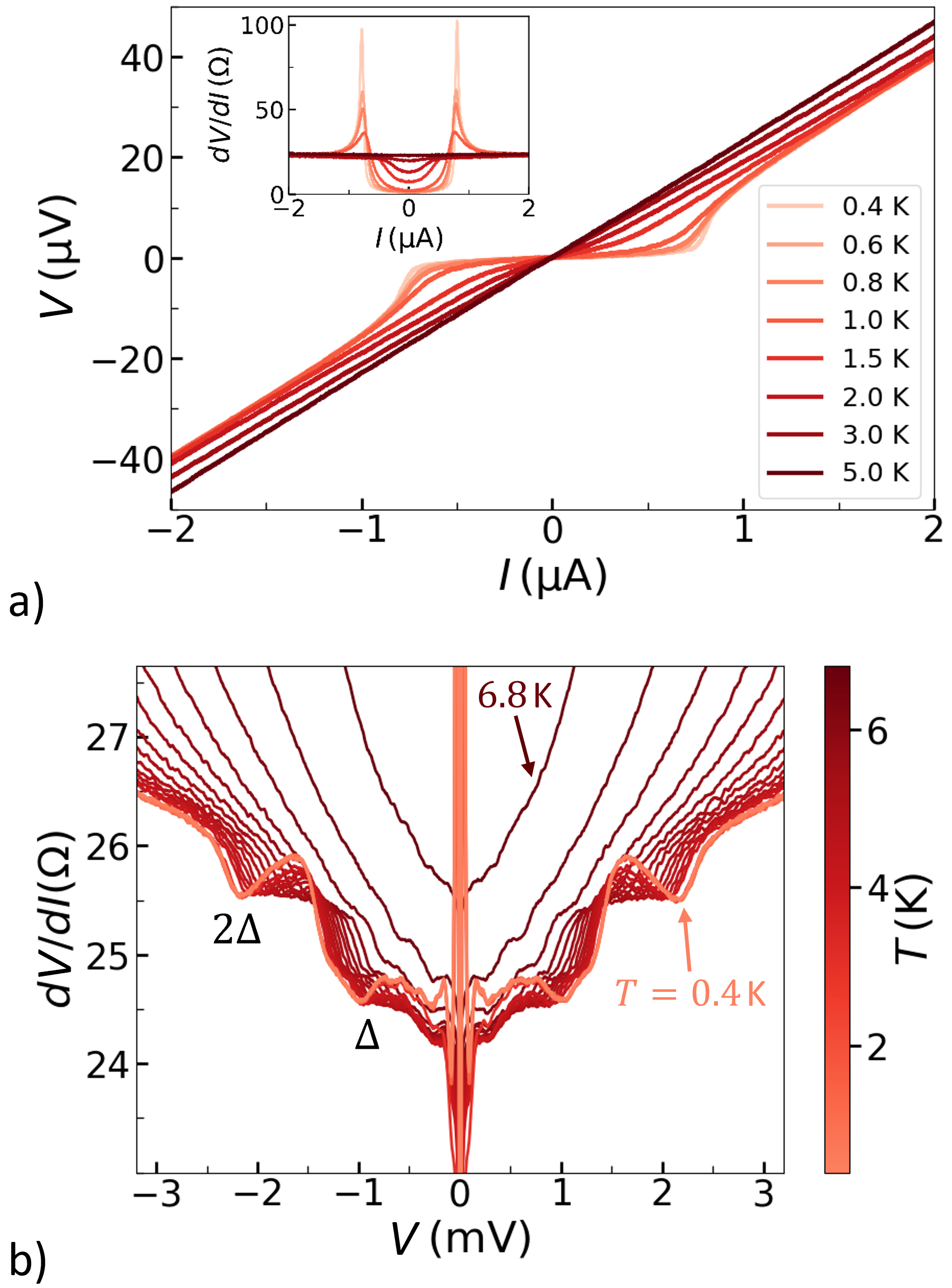}
\caption{a) Current-voltage characteristics of sample A at various temperatures. In the inset the corresponding differential resistance measurement results are shown. b) 
Differential resistance vs. voltage for an increased bias range at temperatures ranging from 0.4\,K to 6.8\,K.} \label{Fig_SQUID_He3_IV_Sam_D}
\end{figure}

First, we focus on the the transport properties of junction A which revealed a clear Josephson junction behavior. The sample comprises a trench width of $300\,$nm and the spacing between the superconducting electrodes is $80\,$nm. At first, the temperature dependence of the induced superconducting transition is investigated. Figure~\ref{Fig_SQUID_He3_IV_Sam_D}a) shows the current-voltage characteristics at various temperatures.  For temperatures up to about 0.8\,K a clear Josephson supercurrent is observed, with a switching current of 795\,nA at 0.4\,K. Up to 2.0\,K the current-voltage characteristics is still non-linear indicating a Josephson supercurrent under the influence of thermal broadening, whereas at larger temperatures a linear behaviour is found. The inset in Fig.~\ref{Fig_SQUID_He3_IV_Sam_D} shows the corresponding differential resistance $dV/dI$ as a function of a DC bias current recorded with an AC current of $20\,$nA. Furthermore, from the linear behaviour far beyond the superconducting regime an excess current of $I_\text{exc}=352\,$nA is extracted. Based on the Blonder-Tinkham-Klapwijk model~\cite{blonder1982transition} a junction transparency of $\tau=0.40$ is derived~\cite{octavio1983subharmonic,flensberg1988subharmonic,niebler2009analytical}. In comparison to the levels of transparency found in literature, this value is compatible and thus suggests a sufficient yet still improvable quality of the junction~\cite{endres2022transparent,galletti2014influence,kunakova2019high,ghatak2018anomalous}. Moreover, the $I_c R_N$ product results in $19.4\,\upmu$V which is a lower boundary for the induced superconducting energy gap.

When increasing the DC bias, a non-trivial subgap structure is observed. Figure~\ref{Fig_SQUID_He3_IV_Sam_D}b) shows the differential resistance measurement results of multiple Andreev reflections. From the temperature dependence of the $2\Delta$ and $\Delta$ feature, a critical temperature of $T_c=7.22\,$K and $\Delta=1.2\,$meV are calculated for the Nb electrodes.

\begin{figure}[hbtp]
\centering
\includegraphics[width=0.47\textwidth]{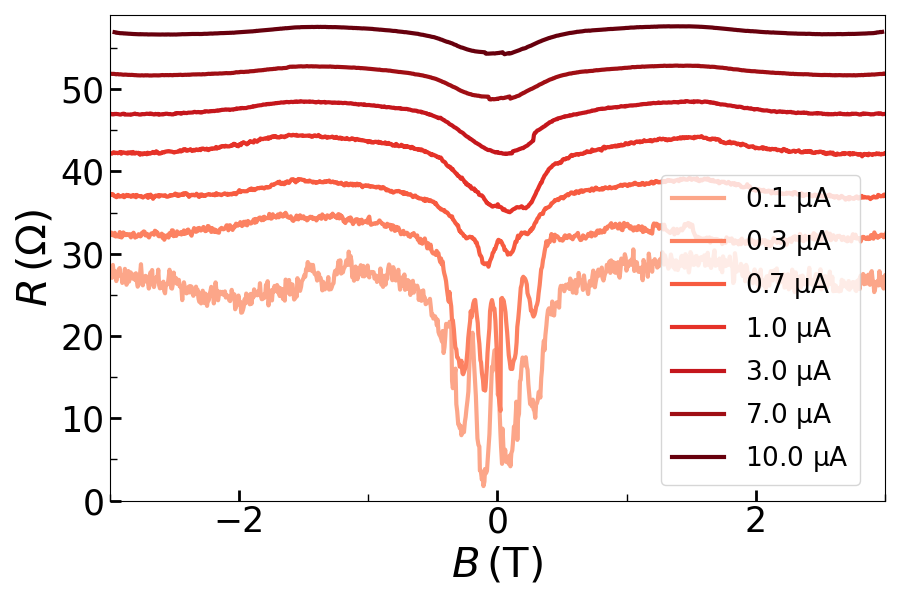}
\caption{Resistance as a function of magnetic field in sample A. For low DC bias currents pronounced SQUID oscillations can be seen that decay with increasing current. The curves are shifted by $5\,\Omega$ each, with the curve for $0.1\,\upmu$A as the reference curve.}\label{Fig_SQUID_He3_Nb_Oscillations}
\end{figure}

Next, the response of the sample to a magnetic field along the nanoribbon axis is investigated. Figure~\ref{Fig_SQUID_He3_Nb_Oscillations} shows the resistance with respect to the magnetic field for different bias currents. For low DC bias currents one can see a sharp dip. Pronounced oscillations are found that decrease with increasing DC bias current.
For the measurement at a DC bias current of $I=0.1\,\upmu$A the extracted oscillation period $\Delta B_\mathrm{exp}$ is $205\,$mT with an according frequency of $f_B \approx 4.9\,$T$^{-1}$. Assuming an $h/2e$ periodicity, the oscillation fits to an area that is about $35\,$\% larger than the expected cross-sectional area. This is explained with a deviation in the thickness of the device, as well as an imperfect trench width caused by uncertainties in the resist process. Thus, these measurements show clear indications of the aforementioned SQUID oscillations where one of weak-link Josephson junctions is formed by the upper topologically protected surface channel while the other one is formed by the corresponding bottom channel (cf. Fig.~\ref{Fig_TI_SQUID_Schema_SEM_TEM}a). The large amplitude at low bias currents and magnetic field is attributed to a dominant transport in the surface states while the bulk remains insulating. This is caused by a lower superconducting coherence lengths in the bulk than the width of the superconducting contacts, allowing only phase-coherent transport on the perimeter of the nanoribbon in the superconducting regime~\cite{schuffelgen2019selective}. The asymmetrical behavior of these oscillations is attributed to sporadically occurring flux trapping at local vortices due to the type-II behavior of the Nb superconductor. Besides the dominant oscillations around zero magnetic field smaller amplitude oscillations are distinguished for $I=0.1\,\upmu$A from the background outside of the superconducting region of the junction, i.e. between $-2$\,T and $-1$\,T. This type of oscillations is also observed for the non-superconducting interferometers, discussed in the following section. 

\section{Non-superconducting Interferometers}
In this section the response to an axial magnetic field $B$ of the non-superconducting samples (B-D) is discussed. The shadow masks of sample B and C have the same dimensions, so the distance between the superconducting electrodes is the same. For those two devices, however, the width of the TI nanoribbon is different so that oscillations with a different frequency are expected. For sample C and D it is vice versa: The TI nanoribbon has the same width but the distance between the Nb electrodes is different. 
\begin{figure}[btp]
\centering
\includegraphics[width=0.47\textwidth]{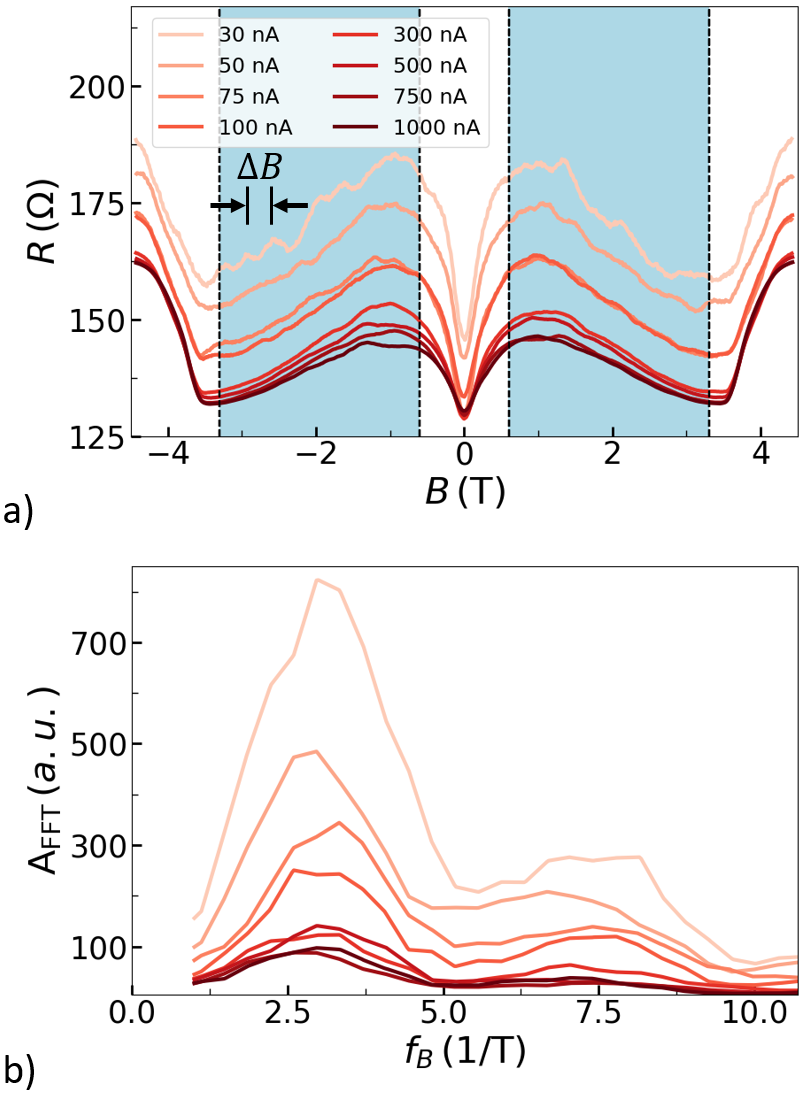}
\caption{a) The resistance $R$ of sample B for different applied DC currents is shown as a function of magnetic field. The colored regions indicate the data that is used for further analysis. The oscillation period $\Delta B$ is indicated in the plot. b) Fourier transformations performed for all applied currents individually reveal the contributing frequencies. The legend is shown in a).}\label{Fig_SQUID_He3_Nonmagn_Sam_B}
\end{figure}

Figure~\ref{Fig_SQUID_He3_Nonmagn_Sam_B}a) shows the magnetoresistance data for sample B recorded with different applied DC bias currents. The measurement data for samples C and D are presented in the Supplementary Material. In addition to the slowly varying background signal, small oscillations are resolved that decrease with increasing bias current. The background is caused by non-coherent trajectories of arbitrary shapes within the bulk of the material. There is also a distinct dip around zero magnetic field. This could be caused by two possible effects. As the most likely explanation, this dip can be due to segments of the wire near the electrodes becoming induced superconducting. As can be seen in Supplementary Figure S5, the dip feature disappears at a temperature of about 4\,K, which makes this explanation plausible. On the other hand, the decrease in resistance could be attributed to the weak antilocalization effect acting on the normally conducting TI. However, the relatively fast disappearance of the dip feature with temperature makes this option less likely. In addition, an increase in resistance around $B=\pm 4\,$T is observed, which is attributed to the switch to the normal conductive state of the superconducting leads. The regions around the dip and the switching of the superconductor are excluded for further analysis, as indicated by the vertical lines in Fig.~\ref{Fig_SQUID_He3_Nonmagn_Sam_B}a).

In order to extract the period of the small oscillations the data for positive and negative magnetic fields are handled separately and averaged in the end. After subtracting the slowly varying background signal, the data is fast Fourier transformed (FFT). The result of the FFT is shown in Fig.~\ref{Fig_SQUID_He3_Nonmagn_Sam_B}b), where a prominent peak at $f_B=2.9\,$T$^{-1}$ corresponding to a period of $\Delta B=345\,$mT is identified. A similar result is observed for positive magnetic fields. The peak position is reproducible among the different bias currents while the amplitude decreases. This is also observed in the raw data in Fig.~\ref{Fig_SQUID_He3_Nonmagn_Sam_B}a) and is therefore consistent and explained by heating effects. When assuming a rectangular cross-section $S$ of the TI nanoribbon using $S\cdot \Delta B = \cdot\phi_{0}$ an accordance of the measured oscillation and the dimensions of the sample is found with a slight deviation of $17\,$\%. 

The oscillations are attributed to phase-coherent Andreev reflections at the interface of the TI and the superconductor, similar to a voltage driven SQUID, as illustrated in the inset of Fig.~\ref{Fig_TI_SQUID_Schema_SEM_TEM}c)~\cite{van1992excess}. Thus, the periodicity corresponds to the superconducting flux quantum $\phi_{0}$. Although the device is in a non-superconducting state, the oscillations nevertheless originate from the presence of superconducting electrodes via coherent Andreev reflection processes mediated by the topological surface states in the lower and upper channel of the nanoribbon. A similar behaviour was observed in metallic ring structures as well as in GaAs/InAs core/shell nanowire in contact with superconducting electrodes \cite{petrashov1993,gul2014flux,haas2018quantum}. 

The same process, as described above, is repeated with sample C and D resulting in resonant Andreev reflections arising from the surface with deviations of $8\,$\% and $12\,$\%, respectively, as summarized in Table~\ref{tab_Nb_SQUIDs}. The deviations can be explained by individual variations in the cross-sectional area due to the under-etch of the SiO$_2$ (see Fig.~\ref{Fig_TI_SQUID_Schema_SEM_TEM}b)), as well as slight variations in the height of the TI caused by the different shapes of the shadow masks. Besides that, deviations arising from the spatial extend of the surface states may occur.

\section{Conclusion} 
In this article the interference of in-situ fabricated axial DC SQUIDs formed by the surface states of a TI nanoribbon was reported. The structure was fabricated by means of selective-area growth and stencil lithography so that the magnetic flux penetrated area of the SQUID is given by the cross-section of the TI nanoribbon. In both, superconducting and non-superconducting devices $\phi_0$-periodic interference mediated by the topologically protected surface states was found for different sample dimensions. In non-superconducting SQUIDs the oscillation was only a minor contribution to the total resistance. The small amplitude was explained with a dominant non-coherent background signal from the bulk of the TI. In a superconducting TI nanoribbon at $T=0.4\,$K a pronounced interference pattern arising from the surface states was observed. The large amplitude at low bias currents and magnetic field is explained by the dominant transport in the surface states while the bulk is not contributing. Hence, the proximity effect represents an important parameter to tune the junction to a surface dominated nano-SQUID.

The presented axial DC SQUID is a promising device towards the detection of Majorana zero modes in TIs, as an efficient exclusion of the bulk contributions is possible only by tuning the phase coherence length in the bulk below and the one of the surface states above the length of the junction. In further experiments for instance tunnel contacts could be used to probe the density of states in the induced superconducting TI in order to detect signatures of Majorana zero modes.

\section{Acknowledgments}
We thank Herbert Kertz for technical assistance, as well as Tobias Schmitt and Kristof Moors for fruitful discussion. All samples have been prepared at the Helmholtz Nano Facility~\cite{albrecht2017hnf}. This work is funded by the Deutsche Forschungsgemeinschaft (DFG, German Research Foundation) under Germany's Excellence Strategy – Cluster of Excellence Matter and Light for Quantum Computing (ML4Q) EXC 2004/1 – 390534769 and by the German Federal Ministry of Education and Research (BMBF) via the Quantum Futur project ‘MajoranaChips’ (Grant No. 13N15264) within the funding program Photonic Research Germany.

%


\clearpage
\widetext

\setcounter{section}{0}
\setcounter{equation}{0}
\setcounter{figure}{0}
\setcounter{table}{0}
\setcounter{page}{1}
\makeatletter
\renewcommand{\thesection}{S\Roman{section}}
\renewcommand{\thesubsection}{\Alph{subsection}}
\renewcommand{\theequation}{S\arabic{equation}}
\renewcommand{\thefigure}{S\arabic{figure}}
\renewcommand{\figurename}{Supplementary Figure}
\renewcommand{\bibnumfmt}[1]{[S#1]}
\renewcommand{\citenumfont}[1]{S#1}

\begin{center}
\textbf{Supplementary Material: Topological insulator based axial superconducting quantum interferometer structures}\end{center}

\section{Fabrication}
\begin{figure}[hbtp]
\centering
\includegraphics[width=\textwidth]{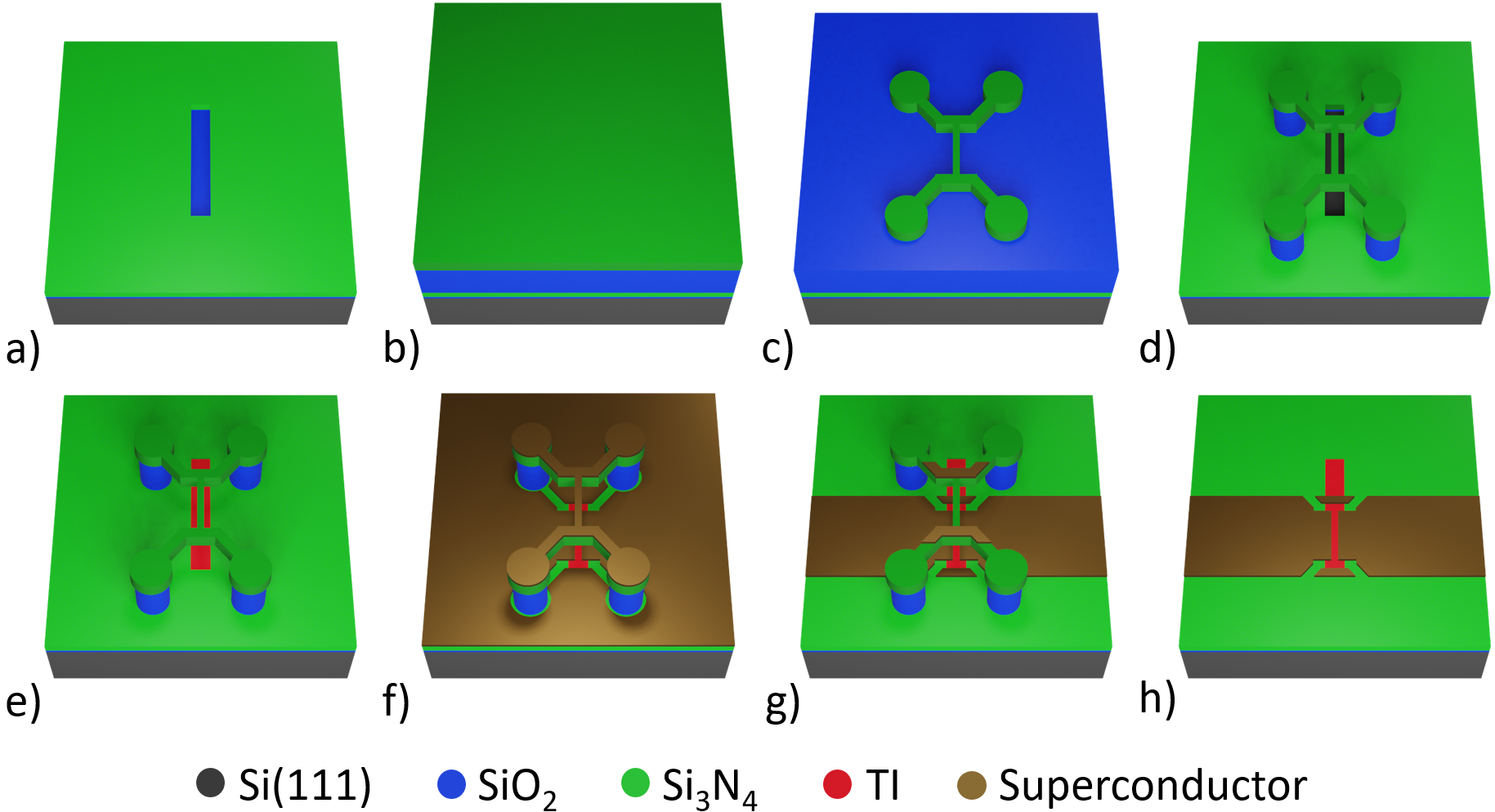}
\caption{Fabrication of axial SQUID devices. a) SiO$_2$ and Si$_3$N$_4$ are deposited on a Si(111) wafer and the topmost layer is etched in a trench shape. b) A second stack of SiO$_2$ and Si$_3$N$_4$ is added for the stencil lithography. Then, the topmost layer is etched into a shadow mask (c) before the SiO$_2$ is removed (d). e) TI is grown selectively in the trench under rotation of the sample. f) The superconductor is deposited in-situ without rotation leaving a free spot under the shadow mask. g) Excess superconductor is removed via RIE to form contact leads. h) After shadow mask removal, the final device consists out of a TI trench contacted with superconductor from both sides on the top.}\label{Supp_Fig_Fabrikation_SQUID_Blender}
\end{figure}

Figure~\ref{Supp_Fig_Fabrikation_SQUID_Blender} depicts schematically the fabrication process of the superconducting quantum interference device (SQUID) described in the main text. Each step is described shortly in the caption of the figure.

\section{Transport characterization on a Hall bar}
\begin{figure}[hbtp]
\centering
\includegraphics[width=0.6\textwidth]{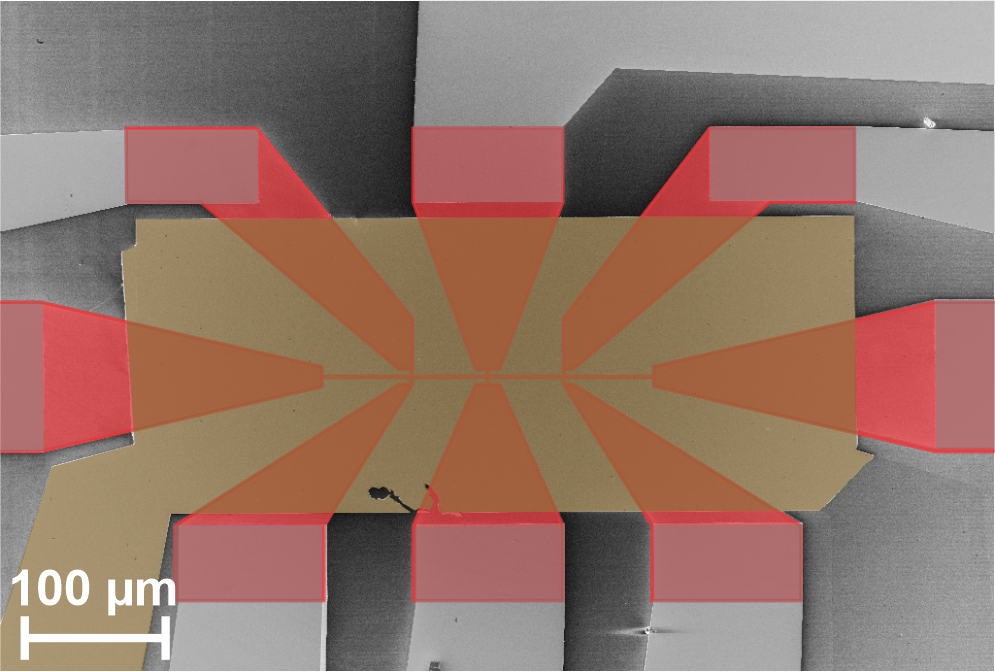}
\caption{SEM image of the ternary micrometer-sized Hall bar. The TI is grown in trenches on a Si(111) surface and contacted far away from the Hall bar with metal contacts. A top gate is placed on the sample, separated by a HfO$_2$ dielectric.}\label{Supp_Fig_Tern_HB}
\end{figure}

A gated reference Hall bar with the same stoichiometry is fabricated selectively in order to gain information about normal transport properties. A false color scanning electron microscopy (SEM) image is shown in Figure~\ref{Supp_Fig_Tern_HB}, where the TI is shown in red and the gate is indicated in brown. The Hall bar has a length of $51\,\upmu$m between each of the side contacts and the width is $1\,\upmu$m.

\begin{figure}[hbtp]
\centering
\includegraphics[width=\textwidth]{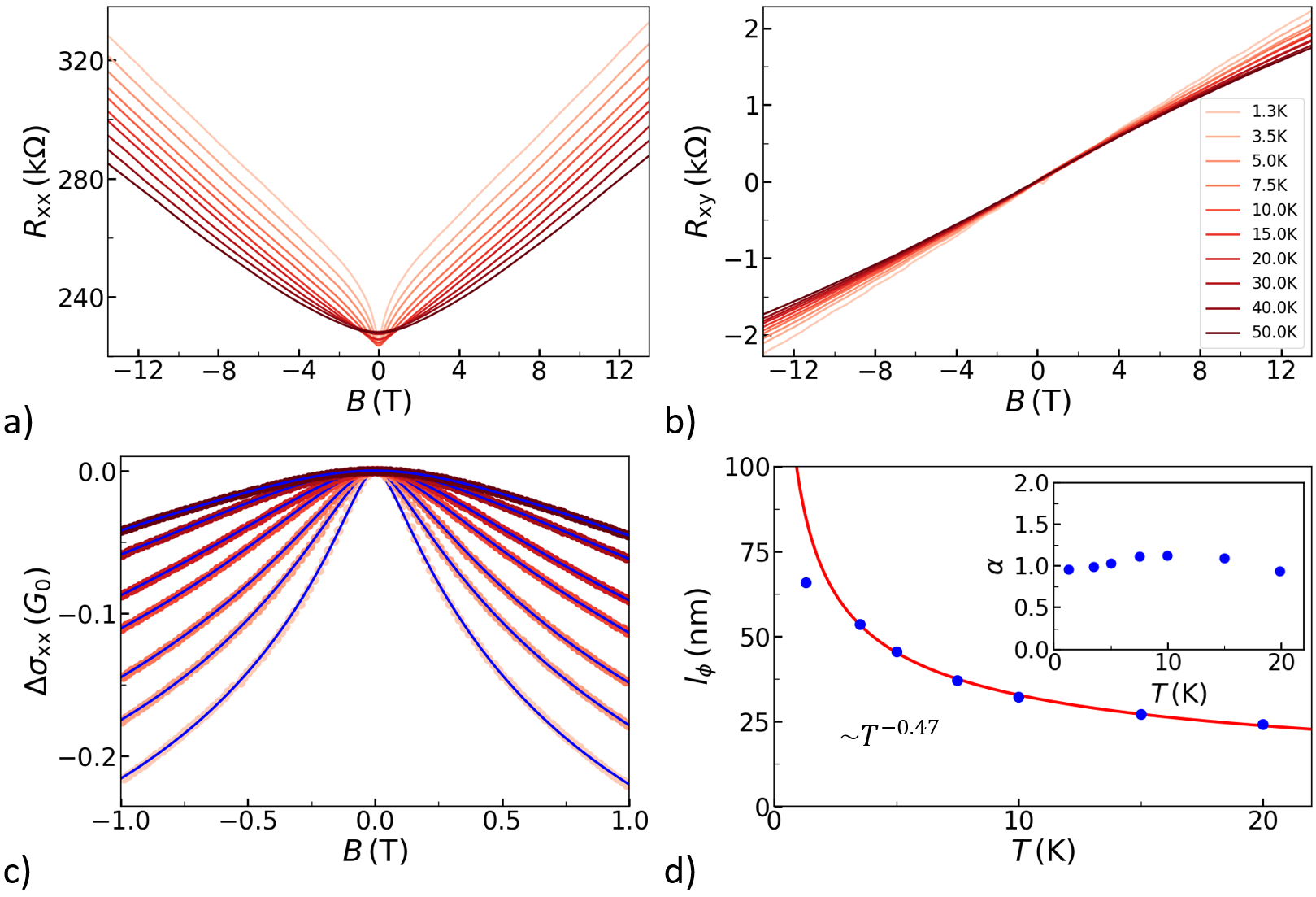}
\caption{Temperature dependent magnetoresistance data. a) The longitudinal resistance $R_{\mathrm{xx}}$ is plotted for temperatures ranging from base temperature to $50\,$K versus the magnetic field $B$. A WAL dip is observed for small magnetic fields and temperatures. b) The corresponding transversal resistance $R_{\mathrm{xy}}$ is shown as a function of the magnetic field for the same temperatures as in a). c) HLN fits are performed for $T\leq20\,$K and plotted in blue. The fit parameters $l_\phi$ and $\alpha$ are plotted in d) and the insert, respectively. The legend is shown in b).}\label{Supp_Fig_MR_Tern_HB_HLN}
\end{figure}

The Hall bar is measured in a variable temperature insert (VTI) cryostat with a base temperature of $1.3\,$K in a four-terminal setup with an AC bias current of $I=100\,$nA at an operating frequency of $31.7\,$Hz. The longitudinal voltage is probed using two neighbouring side contacts while for the Hall voltage opposing contacts are used. The temperature dependent longitudinal magnetoresistance data is shown in Figure~\ref{Supp_Fig_MR_Tern_HB_HLN}a) and the Hall signal is plotted in b). Using a classical Hall approach and the Drude model, a 2D charge carrier concentration of $n_{\text{2D}}=3.72\cdot 10^{12}\,$cm$^{-2}$ and a mobility of $\mu=380\,$cm$^2$/Vs are derived at base temperature.

For temperatures up to $20\,$K a weak antilocalization (WAL) dip can be distinguished from the parabolic background in Figure~\ref{Supp_Fig_MR_Tern_HB_HLN}a). For $B<1\,$T the data is analyzed in more detail following the Hikami, Larkin and Nagaoka (HLN) model~\cite{hikami1980spin}:
\begin{equation}
    \Delta G_{\text{HLN}}=-\alpha \frac{e^2}{2\pi^2\hbar}\left[ \ln\left(\frac{B_\phi}{B}\right)-\psi\left(\frac{1}{2}+\frac{B_\phi}{B}\right)\right]~.\label{eq_HLN}
\end{equation}
Here, $2 |\alpha|$ describes the number of contributing channels and $B_\phi$ is connected to the phase-coherence length $l_\phi$: $B_\phi=\hbar^2(16e^2l_\phi)^{-1}$. The longitudinal conductivity is calculated and the conductivity at zero magnetic field $\sigma_{xx}(B=0\,\text{T})$ is subtracted for each temperature. The difference $\Delta\sigma_{xx}$ is shown in Figure~\ref{Supp_Fig_MR_Tern_HB_HLN}c). For both, positive and negative magnetic fields the HLN formula is fitted. The results for $l_\phi$ and $\alpha$ are averaged for both fits for each temperature and shown in Figure~\ref{Supp_Fig_MR_Tern_HB_HLN}d) and the insert, respectively. $\alpha$ is nearly constant with a value around one for all temperatures indicating transport through two channels. The phase-coherence length is limited to rather small values and decreases with increasing temperature, because the electrons are affected by electron-phonon and electron-electron interactions~\cite{Lin02}. Furthermore, the data is fitted with $l_\phi(T)=a\cdot T^{-b}$. The first data point is not taken into account, because there the HLN fit is least accurate. This may be caused by additional coherent loops in the bulk contributing to the WAL dip at sufficiently low temperature. From the fit an exponent of $0.47$ is deduced which is close to $0.5$ that is expected for 2D transport. Extrapolating the fit to temperatures reached in the $^3$He cryostat an increase of the phase coherence length in the topologically protected surface states to hundreds of nanometers is expected.

\section{TEM Analysis}
\begin{figure}[hbtp]
\centering
\includegraphics[width=0.7\textwidth]{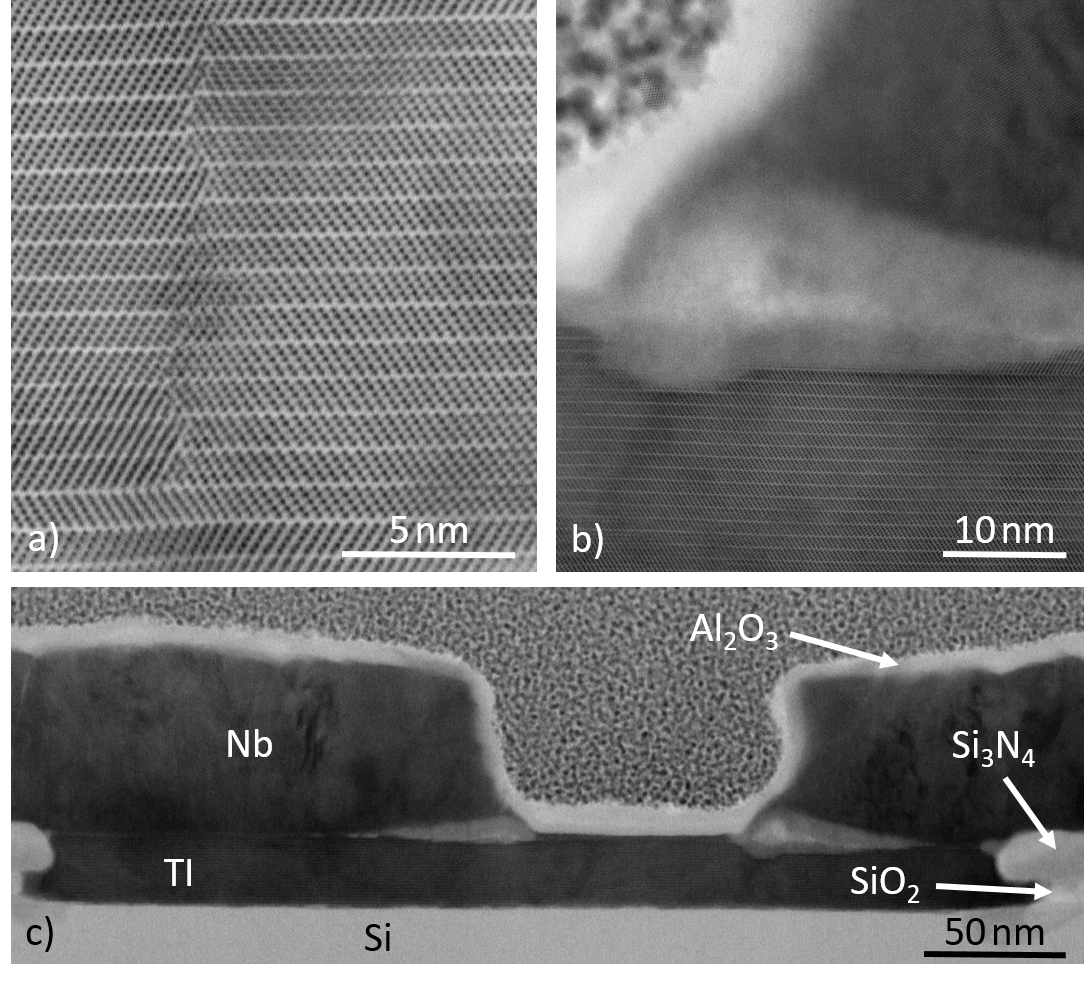}
\caption{Bright-field TEM images of a cut through the sample. a) A zoom into the TI reveals a layering of the atoms in quintuple layers with some twin domains. b) The areas at the edges of the superconductor close to the TI are lifted. c) Overview image showing the complete junction.}\label{Supp_Fig_TEM_Paper}
\end{figure}

The presented transmission electron microscopy (TEM) data is obtained from a reference junction grown on the same chip close to the electrically characterized junctions but split off from the chip before cooldown. The lamella is prepared by cutting the sample from one Nb contact to the other one in the middle of the junction via focused ion beam (FIB). Figure~\ref{Supp_Fig_TEM_Paper}c) shows a bright field TEM image of the junction. In Figure~\ref{Supp_Fig_TEM_Paper}a) a zoom into the TI structure is shown. One can see that the TI is grown in quintuple layers but the zoom-in also reveals the presence of some twin domains. In order to compare the calculated thicknesses of the TI from the oscillations in the magnetotransport measurements to the real thickness, the height of the TI nanoribbon is determined from the images. Here, an average of $24.2\,$nm over the sample is found with a maximum of $25.5\,$nm and a minimum of $22.2\,$nm.

Furthermore, the interface of the TI and the Nb is well defined far away from the junction. However, close to the junction a lighter area is observed. A zoom-in is shown in Figure~\ref{Supp_Fig_TEM_Paper}b). Strain arising from the different lattices of the materials elevated the ends of the superconductor. It remains unclear whether this happened during the preparation of the lamella. Compared to conventional Josephson junctions the contacting from the sides could be a possible reason. Depending on if this partial lift of the electrodes originates in the preparation of the lamella or not, possibly the top Josephson junction is longer than expected. This is not concerned as a disadvantage, as long as a sufficient contact area is left, since this would make the lengths of the two Josephson junctions even more comparable. Furthermore, this effect is also slightly seen at the Si$_3$N$_4$, which reinforces that this lift could already present before the preparation via FIB.

\section{Differential resistance measurements of sample B}
\begin{figure}[btp]
\centering
\includegraphics[width=\textwidth]{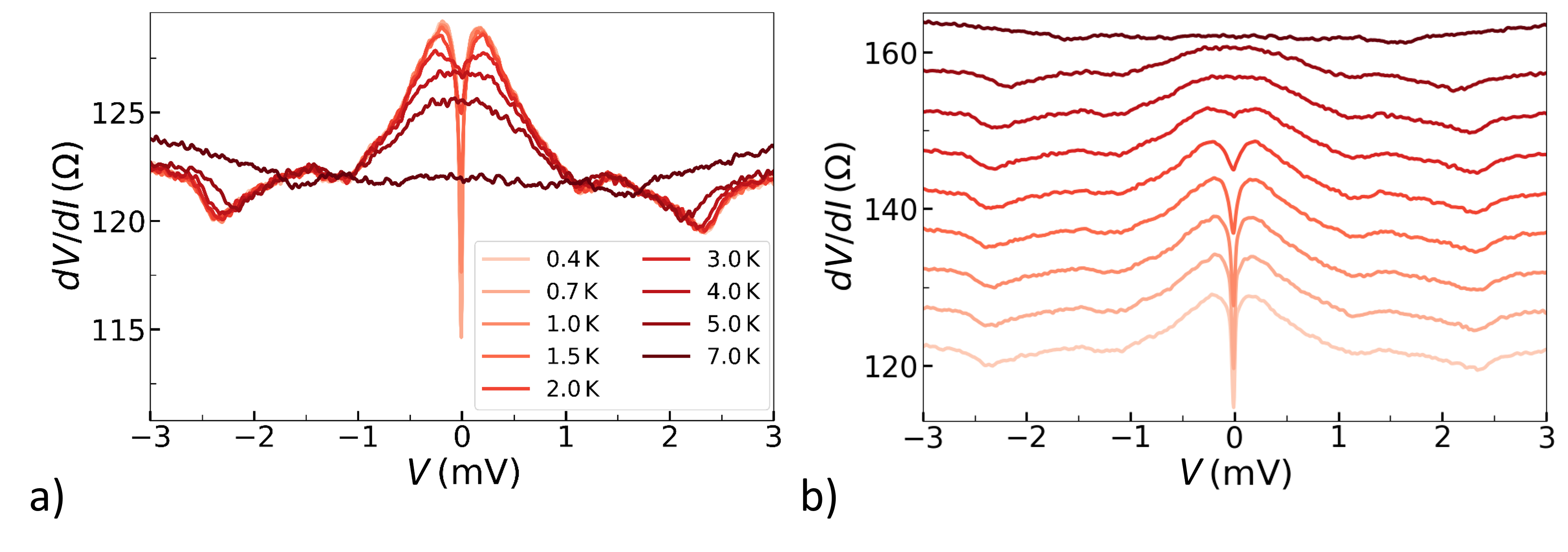}
\caption{Temperature dependent differential resistance of sample A. a) Differential resistance measurements reveal a non-trivial gap structure of the Nb due to multiple Andreev reflections. b) The data shown in a) is plotted with a shift of $5\,\mathrm{\Omega}$ for each temperature. A temperature dependence of the dips can be seen.}\label{Supp_Fig_MAR_Nb_SQUID_Sample_B_paper}
\end{figure}

In Figure~\ref{Supp_Fig_MAR_Nb_SQUID_Sample_B_paper}a) the differential resistance $dV/dI$ is plotted for different temperatures against the DC bias voltage. The drop in resistance at $V=0$ is caused by the proximity effect. As the resistance does not vanish completely, it is concluded that only parts of the sample turn induced superconducting, but there is no superconducting connection between the leads. In Figure~\ref{Supp_Fig_MAR_Nb_SQUID_Sample_B_paper}b) the data shown in Figure~\ref{Supp_Fig_MAR_Nb_SQUID_Sample_B_paper}a) is shifted by $5\,\mathrm{\Omega}$ for each temperature. Here, temperature dependent multiple Andreev reflections are observed.

\section{Background subtraction}
\begin{figure}[btp]
\centering
\includegraphics[width=\textwidth]{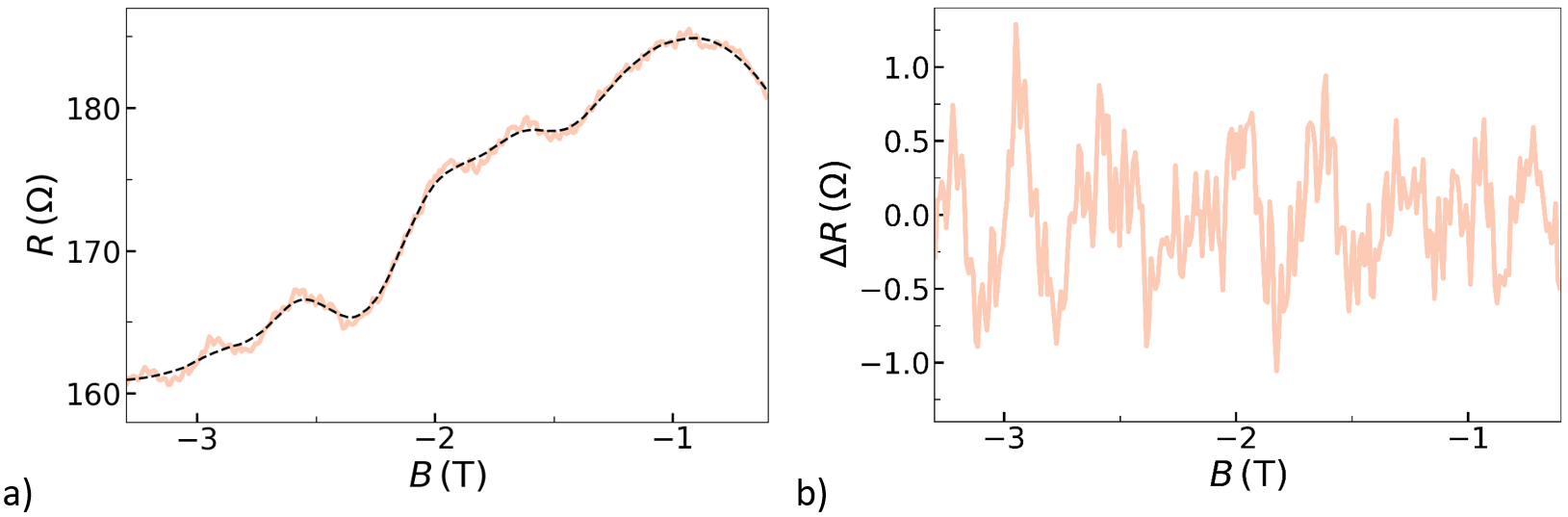}
\caption{Background subtraction of sample B. a) The data for negative magnetic fields for $I=30\,$nA is plotted together with the determined background signal (dotted curve). In b) the difference is shown. Only weak oscillations are observed.}\label{Supp_Fig_SQUID_He3_Background}
\end{figure}

As mentioned in the main text, for the non-superconducting samples a slowly varying background is subtracted from the sample. Figure~\ref{Supp_Fig_SQUID_He3_Background}a) shows exemplarily the signal in the relevant region for negative magnetic fields together with the background that is indicated as a dotted line. Only where the variation of the background signal is moderate, the SQUID oscillation can be seen by eye. The difference of the raw signal and the background is depicted in Figure~\ref{Supp_Fig_SQUID_He3_Background}b). Here, oscillations that are partially just in the order of the noise level are observed. This reinforces the importance of the proximity effect as the oscillation amplitude increases significantly for superconducting samples.
 
\section{Magnetotransport data of sample C and D}
\begin{figure}[btp]
\centering
\includegraphics[width=\textwidth]{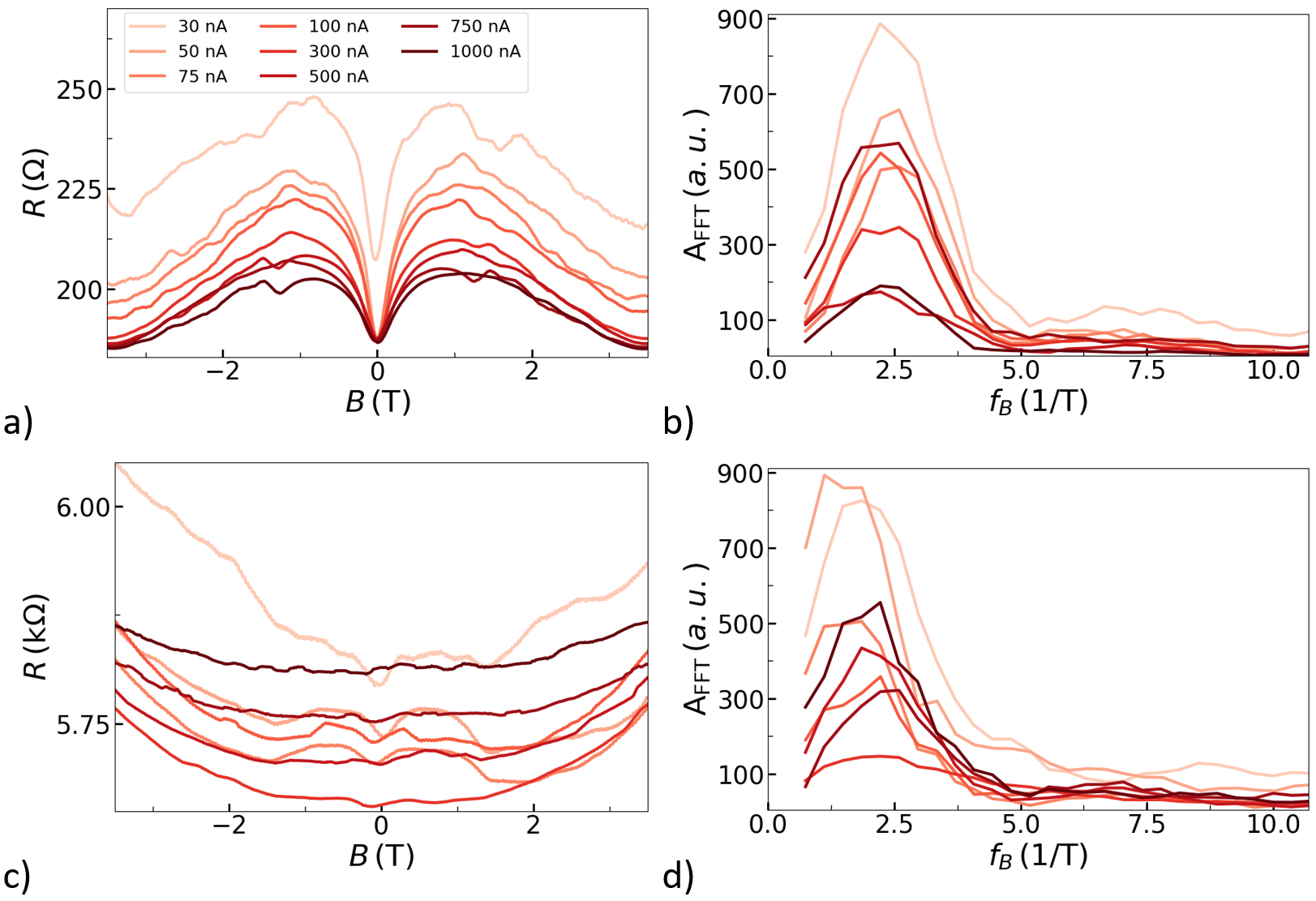}
\caption{The raw data of sample B for different applied currents is shown in a) and the Fourier transformations performed for all applied currents is shown in b). c) and d) respectively show the results for the same measurements of sample C. The legend is shown in a).}\label{Supp_Fig_SQUID_He3_Nonmagn_Sam_AC}
\end{figure}

In Figure~\ref{Supp_Fig_SQUID_He3_Nonmagn_Sam_AC}a) the magnetoresistance data of sample C is shown for different bias currents. Again a dip around zero magnetic field is observed. From the corresponding FFT that is plotted in b) a peak at $f_B=2.2 \,$T$^{-1}$ is identified.\\
The same measurements performed on sample D are shown in Figure~\ref{Supp_Fig_SQUID_He3_Nonmagn_Sam_AC}c) and d), respectively. Here, especially the lowest bias current shows a slight asymmetric behaviour with respect to the applied magnetic field. The impact of the asymmetry can be seen in the Fourier transformation as the peak for $I=50\,$nA is higher than the one at $I=30\,$nA. A peak position of $f_B=2.1\,$T$^{-1}$ is identified. Furthermore, the resistance of the sample is higher which may be caused by a lower mobility due to internal effects. Besides that, the dip around zero magnetic field is less pronounced.

%


\begin{thebibliography}{34}%
\makeatletter
\providecommand \@ifxundefined [1]{%
 \@ifx{#1\undefined}
}%
\providecommand \@ifnum [1]{%
 \ifnum #1\expandafter \@firstoftwo
 \else \expandafter \@secondoftwo
 \fi
}%
\providecommand \@ifx [1]{%
 \ifx #1\expandafter \@firstoftwo
 \else \expandafter \@secondoftwo
 \fi
}%
\providecommand \natexlab [1]{#1}%
\providecommand \enquote  [1]{``#1''}%
\providecommand \bibnamefont  [1]{#1}%
\providecommand \bibfnamefont [1]{#1}%
\providecommand \citenamefont [1]{#1}%
\providecommand \href@noop [0]{\@secondoftwo}%
\providecommand \href [0]{\begingroup \@sanitize@url \@href}%
\providecommand \@href[1]{\@@startlink{#1}\@@href}%
\providecommand \@@href[1]{\endgroup#1\@@endlink}%
\providecommand \@sanitize@url [0]{\catcode `\\12\catcode `\$12\catcode
  `\&12\catcode `\#12\catcode `\^12\catcode `\_12\catcode `\%12\relax}%
\providecommand \@@startlink[1]{}%
\providecommand \@@endlink[0]{}%
\providecommand \url  [0]{\begingroup\@sanitize@url \@url }%
\providecommand \@url [1]{\endgroup\@href {#1}{\urlprefix }}%
\providecommand \urlprefix  [0]{URL }%
\providecommand \Eprint [0]{\href }%
\providecommand \doibase [0]{http://dx.doi.org/}%
\providecommand \selectlanguage [0]{\@gobble}%
\providecommand \bibinfo  [0]{\@secondoftwo}%
\providecommand \bibfield  [0]{\@secondoftwo}%
\providecommand \translation [1]{[#1]}%
\providecommand \BibitemOpen [0]{}%
\providecommand \bibitemStop [0]{}%
\providecommand \bibitemNoStop [0]{.\EOS\space}%
\providecommand \EOS [0]{\spacefactor3000\relax}%
\providecommand \BibitemShut  [1]{\csname bibitem#1\endcsname}%
\let\auto@bib@innerbib\@empty
\bibitem [{\citenamefont {He}\ \emph {et~al.}(2019)\citenamefont {He},
  \citenamefont {Sun},\ and\ \citenamefont {He}}]{he2019topological}%
  \BibitemOpen
  \bibfield  {author} {\bibinfo {author} {\bibfnamefont {M.}~\bibnamefont
  {He}}, \bibinfo {author} {\bibfnamefont {H.}~\bibnamefont {Sun}}, \ and\
  \bibinfo {author} {\bibfnamefont {Q.~L.}\ \bibnamefont {He}},\ }\href
  {\doibase https://doi.org/10.1007/s11467-019-0893-4} {\bibfield  {journal}
  {\bibinfo  {journal} {Frontiers of Physics}\ }\textbf {\bibinfo {volume}
  {14}},\ \bibinfo {pages} {1} (\bibinfo {year} {2019})}\BibitemShut {NoStop}%
\bibitem [{\citenamefont {Breunig}\ and\ \citenamefont
  {Ando}(2022)}]{breunig2022opportunities}%
  \BibitemOpen
  \bibfield  {author} {\bibinfo {author} {\bibfnamefont {O.}~\bibnamefont
  {Breunig}}\ and\ \bibinfo {author} {\bibfnamefont {Y.}~\bibnamefont {Ando}},\
  }\href {\doibase https://doi.org/10.1038/s42254-021-00402-6} {\bibfield
  {journal} {\bibinfo  {journal} {Nature Reviews Physics}\ }\textbf {\bibinfo
  {volume} {4}},\ \bibinfo {pages} {184} (\bibinfo {year} {2022})}\BibitemShut
  {NoStop}%
\bibitem [{\citenamefont {Hasan}\ and\ \citenamefont
  {Kane}(2010)}]{hasan2010colloquium}%
  \BibitemOpen
  \bibfield  {author} {\bibinfo {author} {\bibfnamefont {M.~Z.}\ \bibnamefont
  {Hasan}}\ and\ \bibinfo {author} {\bibfnamefont {C.~L.}\ \bibnamefont
  {Kane}},\ }\href {\doibase https://doi.org/10.1103/RevModPhys.82.3045}
  {\bibfield  {journal} {\bibinfo  {journal} {Reviews of Modern Physics}\
  }\textbf {\bibinfo {volume} {82}},\ \bibinfo {pages} {3045} (\bibinfo {year}
  {2010})}\BibitemShut {NoStop}%
\bibitem [{\citenamefont {Zhao}\ and\ \citenamefont
  {Wang}(2014)}]{zhao2014topological}%
  \BibitemOpen
  \bibfield  {author} {\bibinfo {author} {\bibfnamefont {Y.}~\bibnamefont
  {Zhao}}\ and\ \bibinfo {author} {\bibfnamefont {Z.}~\bibnamefont {Wang}},\
  }\href {\doibase http://dx.doi.org/10.1103/PhysRevB.89.075111} {\bibfield
  {journal} {\bibinfo  {journal} {Physical Review B}\ }\textbf {\bibinfo
  {volume} {89}},\ \bibinfo {pages} {075111} (\bibinfo {year}
  {2014})}\BibitemShut {NoStop}%
\bibitem [{\citenamefont {Ando}(2013)}]{Ando13}%
  \BibitemOpen
  \bibfield  {author} {\bibinfo {author} {\bibfnamefont {Y.}~\bibnamefont
  {Ando}},\ }\href {\doibase 10.7566/JPSJ.82.102001} {\bibfield  {journal}
  {\bibinfo  {journal} {Journal of the Physical Society of Japan}\ }\textbf
  {\bibinfo {volume} {82}},\ \bibinfo {pages} {102001} (\bibinfo {year}
  {2013})}\BibitemShut {NoStop}%
\bibitem [{\citenamefont {Moore}(2010)}]{Moore10}%
  \BibitemOpen
  \bibfield  {author} {\bibinfo {author} {\bibfnamefont {J.}~\bibnamefont
  {Moore}},\ }\href {\doibase doi:10.1038/nature08916} {\bibfield  {journal}
  {\bibinfo  {journal} {Nature}\ }\textbf {\bibinfo {volume} {5}},\ \bibinfo
  {pages} {194} (\bibinfo {year} {2010})}\BibitemShut {NoStop}%
\bibitem [{\citenamefont {Kitaev}(2003)}]{Kitaev03}%
  \BibitemOpen
  \bibfield  {author} {\bibinfo {author} {\bibfnamefont {A.}~\bibnamefont
  {Kitaev}},\ }\href {\doibase https://doi.org/10.1016/S0003-4916(02)00018-0}
  {\bibfield  {journal} {\bibinfo  {journal} {Annals of Physics}\ }\textbf
  {\bibinfo {volume} {303}},\ \bibinfo {pages} {2 } (\bibinfo {year}
  {2003})}\BibitemShut {NoStop}%
\bibitem [{\citenamefont {Nayak}\ \emph {et~al.}(2008)\citenamefont {Nayak},
  \citenamefont {Simon}, \citenamefont {Stern}, \citenamefont {Freedman},\ and\
  \citenamefont {Sarma}}]{Nayak08}%
  \BibitemOpen
  \bibfield  {author} {\bibinfo {author} {\bibfnamefont {C.}~\bibnamefont
  {Nayak}}, \bibinfo {author} {\bibfnamefont {S.~H.}\ \bibnamefont {Simon}},
  \bibinfo {author} {\bibfnamefont {A.}~\bibnamefont {Stern}}, \bibinfo
  {author} {\bibfnamefont {M.}~\bibnamefont {Freedman}}, \ and\ \bibinfo
  {author} {\bibfnamefont {S.~D.}\ \bibnamefont {Sarma}},\ }\href {\doibase
  https://doi.org/10.1103/RevModPhys.80.1083} {\bibfield  {journal} {\bibinfo
  {journal} {Reviews of Modern Physics}\ }\textbf {\bibinfo {volume} {80}},\
  \bibinfo {pages} {1083} (\bibinfo {year} {2008})}\BibitemShut {NoStop}%
\bibitem [{\citenamefont {Hyart}\ \emph {et~al.}(2013)\citenamefont {Hyart},
  \citenamefont {van Heck}, \citenamefont {Fulga}, \citenamefont {Burrello},
  \citenamefont {Akhmerov},\ and\ \citenamefont {Beenakker}}]{Hyart13}%
  \BibitemOpen
  \bibfield  {author} {\bibinfo {author} {\bibfnamefont {T.}~\bibnamefont
  {Hyart}}, \bibinfo {author} {\bibfnamefont {B.}~\bibnamefont {van Heck}},
  \bibinfo {author} {\bibfnamefont {I.~C.}\ \bibnamefont {Fulga}}, \bibinfo
  {author} {\bibfnamefont {M.}~\bibnamefont {Burrello}}, \bibinfo {author}
  {\bibfnamefont {A.~R.}\ \bibnamefont {Akhmerov}}, \ and\ \bibinfo {author}
  {\bibfnamefont {C.~W.~J.}\ \bibnamefont {Beenakker}},\ }\href {\doibase
  10.1103/PhysRevB.88.035121} {\bibfield  {journal} {\bibinfo  {journal} {Phys.
  Rev. B}\ }\textbf {\bibinfo {volume} {88}},\ \bibinfo {pages} {035121}
  (\bibinfo {year} {2013})}\BibitemShut {NoStop}%
\bibitem [{\citenamefont {Alicea}(2012)}]{alicea2012new}%
  \BibitemOpen
  \bibfield  {author} {\bibinfo {author} {\bibfnamefont {J.}~\bibnamefont
  {Alicea}},\ }\href {\doibase https://doi.org/10.1088/0034-4885/75/7/076501}
  {\bibfield  {journal} {\bibinfo  {journal} {Reports on Progress in Physics}\
  }\textbf {\bibinfo {volume} {75}},\ \bibinfo {pages} {076501} (\bibinfo
  {year} {2012})}\BibitemShut {NoStop}%
\bibitem [{\citenamefont {Fu}\ and\ \citenamefont
  {Kane}(2008)}]{fu2008superconducting}%
  \BibitemOpen
  \bibfield  {author} {\bibinfo {author} {\bibfnamefont {L.}~\bibnamefont
  {Fu}}\ and\ \bibinfo {author} {\bibfnamefont {C.~L.}\ \bibnamefont {Kane}},\
  }\href {\doibase https://doi.org/10.1103/PhysRevLett.100.096407} {\bibfield
  {journal} {\bibinfo  {journal} {Physical Review Letters}\ }\textbf {\bibinfo
  {volume} {100}},\ \bibinfo {pages} {096407} (\bibinfo {year}
  {2008})}\BibitemShut {NoStop}%
\bibitem [{\citenamefont {Veldhorst}\ \emph {et~al.}(2012)\citenamefont
  {Veldhorst}, \citenamefont {Molenaar}, \citenamefont {Verwijs}, \citenamefont
  {Hilgenkamp},\ and\ \citenamefont {Brinkman}}]{veldhorst2012optimizing}%
  \BibitemOpen
  \bibfield  {author} {\bibinfo {author} {\bibfnamefont {M.}~\bibnamefont
  {Veldhorst}}, \bibinfo {author} {\bibfnamefont {C.}~\bibnamefont {Molenaar}},
  \bibinfo {author} {\bibfnamefont {C.}~\bibnamefont {Verwijs}}, \bibinfo
  {author} {\bibfnamefont {H.}~\bibnamefont {Hilgenkamp}}, \ and\ \bibinfo
  {author} {\bibfnamefont {A.}~\bibnamefont {Brinkman}},\ }\href {\doibase
  http://dx.doi.org/10.1103/PhysRevB.86.024509} {\bibfield  {journal} {\bibinfo
   {journal} {Physical Review B}\ }\textbf {\bibinfo {volume} {86}},\ \bibinfo
  {pages} {024509} (\bibinfo {year} {2012})}\BibitemShut {NoStop}%
\bibitem [{\citenamefont {Lucignano}\ \emph {et~al.}(2013)\citenamefont
  {Lucignano}, \citenamefont {Tafuri},\ and\ \citenamefont
  {Tagliacozzo}}]{lucignano2013topological}%
  \BibitemOpen
  \bibfield  {author} {\bibinfo {author} {\bibfnamefont {P.}~\bibnamefont
  {Lucignano}}, \bibinfo {author} {\bibfnamefont {F.}~\bibnamefont {Tafuri}}, \
  and\ \bibinfo {author} {\bibfnamefont {A.}~\bibnamefont {Tagliacozzo}},\
  }\href {\doibase http://dx.doi.org/10.1103/PhysRevB.88.184512} {\bibfield
  {journal} {\bibinfo  {journal} {Physical Review B}\ }\textbf {\bibinfo
  {volume} {88}},\ \bibinfo {pages} {184512} (\bibinfo {year}
  {2013})}\BibitemShut {NoStop}%
\bibitem [{\citenamefont {Likharev}(1979)}]{likharev1979}%
  \BibitemOpen
  \bibfield  {author} {\bibinfo {author} {\bibfnamefont {K.~K.}\ \bibnamefont
  {Likharev}},\ }\href {\doibase 10.1103/RevModPhys.51.101} {\bibfield
  {journal} {\bibinfo  {journal} {Rev. Mod. Phys.}\ }\textbf {\bibinfo {volume}
  {51}},\ \bibinfo {pages} {101} (\bibinfo {year} {1979})}\BibitemShut
  {NoStop}%
\bibitem [{\citenamefont {Sch{\"a}pers}(2001)}]{schapers2001superconductor}%
  \BibitemOpen
  \bibfield  {author} {\bibinfo {author} {\bibfnamefont {T.}~\bibnamefont
  {Sch{\"a}pers}},\ }\href {\doibase https://doi.org/10.1007/3-540-45525-6}
  {\emph {\bibinfo {title} {Superconductor/semiconductor junctions}}},\ Vol.\
  \bibinfo {volume} {174}\ (\bibinfo  {publisher} {Springer Science \& Business
  Media},\ \bibinfo {year} {2001})\BibitemShut {NoStop}%
\bibitem [{\citenamefont {Sch{\"u}ffelgen}\ \emph {et~al.}(2019)\citenamefont
  {Sch{\"u}ffelgen}, \citenamefont {Rosenbach}, \citenamefont {Li},
  \citenamefont {Schmitt}, \citenamefont {Schleenvoigt}, \citenamefont {Jalil},
  \citenamefont {Schmitt}, \citenamefont {K{\"o}lzer}, \citenamefont {Wang},
  \citenamefont {Bennemann} \emph {et~al.}}]{schuffelgen2019selective}%
  \BibitemOpen
  \bibfield  {author} {\bibinfo {author} {\bibfnamefont {P.}~\bibnamefont
  {Sch{\"u}ffelgen}}, \bibinfo {author} {\bibfnamefont {D.}~\bibnamefont
  {Rosenbach}}, \bibinfo {author} {\bibfnamefont {C.}~\bibnamefont {Li}},
  \bibinfo {author} {\bibfnamefont {T.~W.}\ \bibnamefont {Schmitt}}, \bibinfo
  {author} {\bibfnamefont {M.}~\bibnamefont {Schleenvoigt}}, \bibinfo {author}
  {\bibfnamefont {A.~R.}\ \bibnamefont {Jalil}}, \bibinfo {author}
  {\bibfnamefont {S.}~\bibnamefont {Schmitt}}, \bibinfo {author} {\bibfnamefont
  {J.}~\bibnamefont {K{\"o}lzer}}, \bibinfo {author} {\bibfnamefont
  {M.}~\bibnamefont {Wang}}, \bibinfo {author} {\bibfnamefont {B.}~\bibnamefont
  {Bennemann}},  \emph {et~al.},\ }\href {\doibase
  https://doi.org/10.1038/s41565-019-0506-y} {\bibfield  {journal} {\bibinfo
  {journal} {Nature Nanotechnology}\ }\textbf {\bibinfo {volume} {14}},\
  \bibinfo {pages} {825} (\bibinfo {year} {2019})}\BibitemShut {NoStop}%
\bibitem [{\citenamefont {Williams}\ \emph {et~al.}(2012)\citenamefont
  {Williams}, \citenamefont {Bestwick}, \citenamefont {Gallagher},
  \citenamefont {Hong}, \citenamefont {Cui}, \citenamefont {Bleich},
  \citenamefont {Analytis}, \citenamefont {Fisher},\ and\ \citenamefont
  {Goldhaber-Gordon}}]{williams2012unconventional}%
  \BibitemOpen
  \bibfield  {author} {\bibinfo {author} {\bibfnamefont {J.}~\bibnamefont
  {Williams}}, \bibinfo {author} {\bibfnamefont {A.}~\bibnamefont {Bestwick}},
  \bibinfo {author} {\bibfnamefont {P.}~\bibnamefont {Gallagher}}, \bibinfo
  {author} {\bibfnamefont {S.~S.}\ \bibnamefont {Hong}}, \bibinfo {author}
  {\bibfnamefont {Y.}~\bibnamefont {Cui}}, \bibinfo {author} {\bibfnamefont
  {A.~S.}\ \bibnamefont {Bleich}}, \bibinfo {author} {\bibfnamefont
  {J.}~\bibnamefont {Analytis}}, \bibinfo {author} {\bibfnamefont
  {I.}~\bibnamefont {Fisher}}, \ and\ \bibinfo {author} {\bibfnamefont
  {D.}~\bibnamefont {Goldhaber-Gordon}},\ }\href {\doibase
  https://doi.org/10.1103/PhysRevLett.109.056803} {\bibfield  {journal}
  {\bibinfo  {journal} {Physical Review Letters}\ }\textbf {\bibinfo {volume}
  {109}},\ \bibinfo {pages} {056803} (\bibinfo {year} {2012})}\BibitemShut
  {NoStop}%
\bibitem [{\citenamefont {Haas}\ \emph {et~al.}(2018)\citenamefont {Haas},
  \citenamefont {Dickheuer}, \citenamefont {Zellekens}, \citenamefont {Rieger},
  \citenamefont {Lepsa}, \citenamefont {L{\"u}th}, \citenamefont
  {Gr{\"u}tzmacher},\ and\ \citenamefont {Sch{\"a}pers}}]{haas2018quantum}%
  \BibitemOpen
  \bibfield  {author} {\bibinfo {author} {\bibfnamefont {F.}~\bibnamefont
  {Haas}}, \bibinfo {author} {\bibfnamefont {S.}~\bibnamefont {Dickheuer}},
  \bibinfo {author} {\bibfnamefont {P.}~\bibnamefont {Zellekens}}, \bibinfo
  {author} {\bibfnamefont {T.}~\bibnamefont {Rieger}}, \bibinfo {author}
  {\bibfnamefont {M.}~\bibnamefont {Lepsa}}, \bibinfo {author} {\bibfnamefont
  {H.}~\bibnamefont {L{\"u}th}}, \bibinfo {author} {\bibfnamefont
  {D.}~\bibnamefont {Gr{\"u}tzmacher}}, \ and\ \bibinfo {author} {\bibfnamefont
  {T.}~\bibnamefont {Sch{\"a}pers}},\ }\href {\doibase
  https://doi.org/10.1088/1361-6641/aabc6d} {\bibfield  {journal} {\bibinfo
  {journal} {Semiconductor Science and Technology}\ }\textbf {\bibinfo {volume}
  {33}},\ \bibinfo {pages} {064001} (\bibinfo {year} {2018})}\BibitemShut
  {NoStop}%
\bibitem [{\citenamefont {Jalil}\ \emph {et~al.}(2023)\citenamefont {Jalil},
  \citenamefont {Sch{\"u}ffelgen}, \citenamefont {Valencia}, \citenamefont
  {Schleenvoigt}, \citenamefont {Ringkamp}, \citenamefont {Mussler},
  \citenamefont {Luysberg}, \citenamefont {Mayer},\ and\ \citenamefont
  {Gr{\"u}tzmacher}}]{jalil2023selective}%
  \BibitemOpen
  \bibfield  {author} {\bibinfo {author} {\bibfnamefont {A.~R.}\ \bibnamefont
  {Jalil}}, \bibinfo {author} {\bibfnamefont {P.}~\bibnamefont
  {Sch{\"u}ffelgen}}, \bibinfo {author} {\bibfnamefont {H.}~\bibnamefont
  {Valencia}}, \bibinfo {author} {\bibfnamefont {M.}~\bibnamefont
  {Schleenvoigt}}, \bibinfo {author} {\bibfnamefont {C.}~\bibnamefont
  {Ringkamp}}, \bibinfo {author} {\bibfnamefont {G.}~\bibnamefont {Mussler}},
  \bibinfo {author} {\bibfnamefont {M.}~\bibnamefont {Luysberg}}, \bibinfo
  {author} {\bibfnamefont {J.}~\bibnamefont {Mayer}}, \ and\ \bibinfo {author}
  {\bibfnamefont {D.}~\bibnamefont {Gr{\"u}tzmacher}},\ }\href {\doibase
  https://doi.org/10.3390/nano13020354} {\bibfield  {journal} {\bibinfo
  {journal} {Nanomaterials}\ }\textbf {\bibinfo {volume} {13}},\ \bibinfo
  {pages} {354} (\bibinfo {year} {2023})}\BibitemShut {NoStop}%
\bibitem [{\citenamefont {Schmitt}\ \emph {et~al.}(2022)\citenamefont
  {Schmitt}, \citenamefont {Connolly}, \citenamefont {Schleenvoigt},
  \citenamefont {Liu}, \citenamefont {Kennedy}, \citenamefont
  {Ch{\'a}vez-Garcia}, \citenamefont {Jalil}, \citenamefont {Bennemann},
  \citenamefont {Trellenkamp}, \citenamefont {Lentz}, \citenamefont {Neumann},
  \citenamefont {Lindstr{\"o}m}, \citenamefont {de~Graaf}, \citenamefont
  {Berenschot}, \citenamefont {Tas}, \citenamefont {Mussler}, \citenamefont
  {Petersson}, \citenamefont {Gr{\"u}tzmacher},\ and\ \citenamefont
  {Sch{\"u}ffelgen}}]{schmitt2022integration}%
  \BibitemOpen
  \bibfield  {author} {\bibinfo {author} {\bibfnamefont {T.~W.}\ \bibnamefont
  {Schmitt}}, \bibinfo {author} {\bibfnamefont {M.~R.}\ \bibnamefont
  {Connolly}}, \bibinfo {author} {\bibfnamefont {M.}~\bibnamefont
  {Schleenvoigt}}, \bibinfo {author} {\bibfnamefont {C.}~\bibnamefont {Liu}},
  \bibinfo {author} {\bibfnamefont {O.}~\bibnamefont {Kennedy}}, \bibinfo
  {author} {\bibfnamefont {J.~M.}\ \bibnamefont {Ch{\'a}vez-Garcia}}, \bibinfo
  {author} {\bibfnamefont {A.~R.}\ \bibnamefont {Jalil}}, \bibinfo {author}
  {\bibfnamefont {B.}~\bibnamefont {Bennemann}}, \bibinfo {author}
  {\bibfnamefont {S.}~\bibnamefont {Trellenkamp}}, \bibinfo {author}
  {\bibfnamefont {F.}~\bibnamefont {Lentz}}, \bibinfo {author} {\bibfnamefont
  {E.}~\bibnamefont {Neumann}}, \bibinfo {author} {\bibfnamefont
  {T.}~\bibnamefont {Lindstr{\"o}m}}, \bibinfo {author} {\bibfnamefont {S.~E.}\
  \bibnamefont {de~Graaf}}, \bibinfo {author} {\bibfnamefont {E.}~\bibnamefont
  {Berenschot}}, \bibinfo {author} {\bibfnamefont {N.}~\bibnamefont {Tas}},
  \bibinfo {author} {\bibfnamefont {G.}~\bibnamefont {Mussler}}, \bibinfo
  {author} {\bibfnamefont {K.~D.}\ \bibnamefont {Petersson}}, \bibinfo {author}
  {\bibfnamefont {D.}~\bibnamefont {Gr{\"u}tzmacher}}, \ and\ \bibinfo {author}
  {\bibfnamefont {P.}~\bibnamefont {Sch{\"u}ffelgen}},\ }\href {\doibase
  https://doi.org/10.1021/acs.nanolett.1c04055} {\bibfield  {journal} {\bibinfo
   {journal} {Nano Letters}\ }\textbf {\bibinfo {volume} {22}},\ \bibinfo
  {pages} {2595} (\bibinfo {year} {2022})}\BibitemShut {NoStop}%
\bibitem [{\citenamefont {K{\"o}lzer}\ \emph {et~al.}(2023)\citenamefont
  {K{\"o}lzer}, \citenamefont {Jalil}, \citenamefont {Rosenbach}, \citenamefont
  {Arndt}, \citenamefont {Mussler}, \citenamefont {Sch{\"u}ffelgen},
  \citenamefont {Gr{\"u}tzmacher}, \citenamefont {L{\"u}th},\ and\
  \citenamefont {Sch{\"a}pers}}]{kolzer2023supercurrent}%
  \BibitemOpen
  \bibfield  {author} {\bibinfo {author} {\bibfnamefont {J.}~\bibnamefont
  {K{\"o}lzer}}, \bibinfo {author} {\bibfnamefont {A.~R.}\ \bibnamefont
  {Jalil}}, \bibinfo {author} {\bibfnamefont {D.}~\bibnamefont {Rosenbach}},
  \bibinfo {author} {\bibfnamefont {L.}~\bibnamefont {Arndt}}, \bibinfo
  {author} {\bibfnamefont {G.}~\bibnamefont {Mussler}}, \bibinfo {author}
  {\bibfnamefont {P.}~\bibnamefont {Sch{\"u}ffelgen}}, \bibinfo {author}
  {\bibfnamefont {D.}~\bibnamefont {Gr{\"u}tzmacher}}, \bibinfo {author}
  {\bibfnamefont {H.}~\bibnamefont {L{\"u}th}}, \ and\ \bibinfo {author}
  {\bibfnamefont {T.}~\bibnamefont {Sch{\"a}pers}},\ }\href {\doibase
  https://doi.org/10.3390/nano13020293} {\bibfield  {journal} {\bibinfo
  {journal} {Nanomaterials}\ }\textbf {\bibinfo {volume} {13}},\ \bibinfo
  {pages} {293} (\bibinfo {year} {2023})}\BibitemShut {NoStop}%
\bibitem [{\citenamefont {Weyrich}\ \emph {et~al.}(2016)\citenamefont
  {Weyrich}, \citenamefont {Dr\"ogeler}, \citenamefont {Kampmeier},
  \citenamefont {Eschbach}, \citenamefont {Mussler}, \citenamefont {Merzenich},
  \citenamefont {Stoica}, \citenamefont {Batov}, \citenamefont {Schubert},
  \citenamefont {Plucinski}, \citenamefont {Beschoten}, \citenamefont
  {Schneider}, \citenamefont {Stampfer}, \citenamefont {Gr\"utzmacher},\ and\
  \citenamefont {Sch\"apers}}]{weyrich2016}%
  \BibitemOpen
  \bibfield  {author} {\bibinfo {author} {\bibfnamefont {C.}~\bibnamefont
  {Weyrich}}, \bibinfo {author} {\bibfnamefont {M.}~\bibnamefont {Dr\"ogeler}},
  \bibinfo {author} {\bibfnamefont {J.}~\bibnamefont {Kampmeier}}, \bibinfo
  {author} {\bibfnamefont {M.}~\bibnamefont {Eschbach}}, \bibinfo {author}
  {\bibfnamefont {G.}~\bibnamefont {Mussler}}, \bibinfo {author} {\bibfnamefont
  {T.}~\bibnamefont {Merzenich}}, \bibinfo {author} {\bibfnamefont
  {T.}~\bibnamefont {Stoica}}, \bibinfo {author} {\bibfnamefont {I.~E.}\
  \bibnamefont {Batov}}, \bibinfo {author} {\bibfnamefont {J.}~\bibnamefont
  {Schubert}}, \bibinfo {author} {\bibfnamefont {L.}~\bibnamefont {Plucinski}},
  \bibinfo {author} {\bibfnamefont {B.}~\bibnamefont {Beschoten}}, \bibinfo
  {author} {\bibfnamefont {C.~M.}\ \bibnamefont {Schneider}}, \bibinfo {author}
  {\bibfnamefont {C.}~\bibnamefont {Stampfer}}, \bibinfo {author}
  {\bibfnamefont {D.}~\bibnamefont {Gr\"utzmacher}}, \ and\ \bibinfo {author}
  {\bibfnamefont {T.}~\bibnamefont {Sch\"apers}},\ }\href {\doibase
  https://doi.org/10.1088/0953-8984/28/49/495501} {\bibfield  {journal}
  {\bibinfo  {journal} {Journal of Physics: Condensed Matter}\ }\textbf
  {\bibinfo {volume} {28}},\ \bibinfo {pages} {495501} (\bibinfo {year}
  {2016})}\BibitemShut {NoStop}%
\bibitem [{\citenamefont {Blonder}\ \emph {et~al.}(1982)\citenamefont
  {Blonder}, \citenamefont {Tinkham},\ and\ \citenamefont
  {Klapwijk}}]{blonder1982transition}%
  \BibitemOpen
  \bibfield  {author} {\bibinfo {author} {\bibfnamefont {G.}~\bibnamefont
  {Blonder}}, \bibinfo {author} {\bibfnamefont {M.}~\bibnamefont {Tinkham}}, \
  and\ \bibinfo {author} {\bibfnamefont {T.}~\bibnamefont {Klapwijk}},\ }\href
  {\doibase https://doi.org/10.1103/PhysRevB.25.4515} {\bibfield  {journal}
  {\bibinfo  {journal} {Physical Review B}\ }\textbf {\bibinfo {volume} {25}},\
  \bibinfo {pages} {4515} (\bibinfo {year} {1982})}\BibitemShut {NoStop}%
\bibitem [{\citenamefont {Octavio}\ \emph {et~al.}(1983)\citenamefont
  {Octavio}, \citenamefont {Tinkham}, \citenamefont {Blonder},\ and\
  \citenamefont {Klapwijk}}]{octavio1983subharmonic}%
  \BibitemOpen
  \bibfield  {author} {\bibinfo {author} {\bibfnamefont {M.}~\bibnamefont
  {Octavio}}, \bibinfo {author} {\bibfnamefont {M.}~\bibnamefont {Tinkham}},
  \bibinfo {author} {\bibfnamefont {G.}~\bibnamefont {Blonder}}, \ and\
  \bibinfo {author} {\bibfnamefont {T.}~\bibnamefont {Klapwijk}},\ }\href
  {\doibase https://doi.org/10.1103/PhysRevB.27.6739} {\bibfield  {journal}
  {\bibinfo  {journal} {Physical Review B}\ }\textbf {\bibinfo {volume} {27}},\
  \bibinfo {pages} {6739} (\bibinfo {year} {1983})}\BibitemShut {NoStop}%
\bibitem [{\citenamefont {Flensberg}\ \emph {et~al.}(1988)\citenamefont
  {Flensberg}, \citenamefont {Hansen},\ and\ \citenamefont
  {Octavio}}]{flensberg1988subharmonic}%
  \BibitemOpen
  \bibfield  {author} {\bibinfo {author} {\bibfnamefont {K.}~\bibnamefont
  {Flensberg}}, \bibinfo {author} {\bibfnamefont {J.~B.}\ \bibnamefont
  {Hansen}}, \ and\ \bibinfo {author} {\bibfnamefont {M.}~\bibnamefont
  {Octavio}},\ }\href {\doibase https://doi.org/10.1103/PhysRevB.38.8707}
  {\bibfield  {journal} {\bibinfo  {journal} {Physical Review B}\ }\textbf
  {\bibinfo {volume} {38}},\ \bibinfo {pages} {8707} (\bibinfo {year}
  {1988})}\BibitemShut {NoStop}%
\bibitem [{\citenamefont {Niebler}\ \emph {et~al.}(2009)\citenamefont
  {Niebler}, \citenamefont {Cuniberti},\ and\ \citenamefont
  {Novotn{\`y}}}]{niebler2009analytical}%
  \BibitemOpen
  \bibfield  {author} {\bibinfo {author} {\bibfnamefont {G.}~\bibnamefont
  {Niebler}}, \bibinfo {author} {\bibfnamefont {G.}~\bibnamefont {Cuniberti}},
  \ and\ \bibinfo {author} {\bibfnamefont {T.}~\bibnamefont {Novotn{\`y}}},\
  }\href {\doibase https://doi.org/10.1088/0953-2048/22/8/085016} {\bibfield
  {journal} {\bibinfo  {journal} {Superconductor Science and Technology}\
  }\textbf {\bibinfo {volume} {22}},\ \bibinfo {pages} {085016} (\bibinfo
  {year} {2009})}\BibitemShut {NoStop}%
\bibitem [{\citenamefont {Endres}\ \emph {et~al.}(2022)\citenamefont {Endres},
  \citenamefont {Kononov}, \citenamefont {Stiefel}, \citenamefont {Wyss},
  \citenamefont {Arachchige}, \citenamefont {Yan}, \citenamefont {Mandrus},
  \citenamefont {Watanabe}, \citenamefont {Taniguchi},\ and\ \citenamefont
  {Sch{\"o}nenberger}}]{endres2022transparent}%
  \BibitemOpen
  \bibfield  {author} {\bibinfo {author} {\bibfnamefont {M.}~\bibnamefont
  {Endres}}, \bibinfo {author} {\bibfnamefont {A.}~\bibnamefont {Kononov}},
  \bibinfo {author} {\bibfnamefont {M.}~\bibnamefont {Stiefel}}, \bibinfo
  {author} {\bibfnamefont {M.}~\bibnamefont {Wyss}}, \bibinfo {author}
  {\bibfnamefont {H.~S.}\ \bibnamefont {Arachchige}}, \bibinfo {author}
  {\bibfnamefont {J.}~\bibnamefont {Yan}}, \bibinfo {author} {\bibfnamefont
  {D.}~\bibnamefont {Mandrus}}, \bibinfo {author} {\bibfnamefont
  {K.}~\bibnamefont {Watanabe}}, \bibinfo {author} {\bibfnamefont
  {T.}~\bibnamefont {Taniguchi}}, \ and\ \bibinfo {author} {\bibfnamefont
  {C.}~\bibnamefont {Sch{\"o}nenberger}},\ }\href {\doibase
  https://doi.org/10.1103/PhysRevMaterials.6.L081201} {\bibfield  {journal}
  {\bibinfo  {journal} {Physical Review Materials}\ }\textbf {\bibinfo {volume}
  {6}},\ \bibinfo {pages} {L081201} (\bibinfo {year} {2022})}\BibitemShut
  {NoStop}%
\bibitem [{\citenamefont {Galletti}\ \emph {et~al.}(2014)\citenamefont
  {Galletti}, \citenamefont {Charpentier}, \citenamefont {Iavarone},
  \citenamefont {Lucignano}, \citenamefont {Massarotti}, \citenamefont
  {Arpaia}, \citenamefont {Suzuki}, \citenamefont {Kadowaki}, \citenamefont
  {Bauch}, \citenamefont {Tagliacozzo} \emph {et~al.}}]{galletti2014influence}%
  \BibitemOpen
  \bibfield  {author} {\bibinfo {author} {\bibfnamefont {L.}~\bibnamefont
  {Galletti}}, \bibinfo {author} {\bibfnamefont {S.}~\bibnamefont
  {Charpentier}}, \bibinfo {author} {\bibfnamefont {M.}~\bibnamefont
  {Iavarone}}, \bibinfo {author} {\bibfnamefont {P.}~\bibnamefont {Lucignano}},
  \bibinfo {author} {\bibfnamefont {D.}~\bibnamefont {Massarotti}}, \bibinfo
  {author} {\bibfnamefont {R.}~\bibnamefont {Arpaia}}, \bibinfo {author}
  {\bibfnamefont {Y.}~\bibnamefont {Suzuki}}, \bibinfo {author} {\bibfnamefont
  {K.}~\bibnamefont {Kadowaki}}, \bibinfo {author} {\bibfnamefont
  {T.}~\bibnamefont {Bauch}}, \bibinfo {author} {\bibfnamefont
  {A.}~\bibnamefont {Tagliacozzo}},  \emph {et~al.},\ }\href {\doibase
  http://dx.doi.org/10.1103/PhysRevB.89.134512} {\bibfield  {journal} {\bibinfo
   {journal} {Physical Review B}\ }\textbf {\bibinfo {volume} {89}},\ \bibinfo
  {pages} {134512} (\bibinfo {year} {2014})}\BibitemShut {NoStop}%
\bibitem [{\citenamefont {Kunakova}\ \emph {et~al.}(2019)\citenamefont
  {Kunakova}, \citenamefont {Bauch}, \citenamefont {Trabaldo}, \citenamefont
  {Andzane}, \citenamefont {Erts},\ and\ \citenamefont
  {Lombardi}}]{kunakova2019high}%
  \BibitemOpen
  \bibfield  {author} {\bibinfo {author} {\bibfnamefont {G.}~\bibnamefont
  {Kunakova}}, \bibinfo {author} {\bibfnamefont {T.}~\bibnamefont {Bauch}},
  \bibinfo {author} {\bibfnamefont {E.}~\bibnamefont {Trabaldo}}, \bibinfo
  {author} {\bibfnamefont {J.}~\bibnamefont {Andzane}}, \bibinfo {author}
  {\bibfnamefont {D.}~\bibnamefont {Erts}}, \ and\ \bibinfo {author}
  {\bibfnamefont {F.}~\bibnamefont {Lombardi}},\ }\href {\doibase
  https://doi.org/10.1063/1.5123554} {\bibfield  {journal} {\bibinfo  {journal}
  {Applied Physics Letters}\ }\textbf {\bibinfo {volume} {115}},\ \bibinfo
  {pages} {172601} (\bibinfo {year} {2019})}\BibitemShut {NoStop}%
\bibitem [{\citenamefont {Ghatak}\ \emph {et~al.}(2018)\citenamefont {Ghatak},
  \citenamefont {Breunig}, \citenamefont {Yang}, \citenamefont {Wang},
  \citenamefont {Taskin},\ and\ \citenamefont {Ando}}]{ghatak2018anomalous}%
  \BibitemOpen
  \bibfield  {author} {\bibinfo {author} {\bibfnamefont {S.}~\bibnamefont
  {Ghatak}}, \bibinfo {author} {\bibfnamefont {O.}~\bibnamefont {Breunig}},
  \bibinfo {author} {\bibfnamefont {F.}~\bibnamefont {Yang}}, \bibinfo {author}
  {\bibfnamefont {Z.}~\bibnamefont {Wang}}, \bibinfo {author} {\bibfnamefont
  {A.~A.}\ \bibnamefont {Taskin}}, \ and\ \bibinfo {author} {\bibfnamefont
  {Y.}~\bibnamefont {Ando}},\ }\href {\doibase
  http://dx.doi.org/10.1021/acs.nanolett.8b02029} {\bibfield  {journal}
  {\bibinfo  {journal} {Nano Letters}\ }\textbf {\bibinfo {volume} {18}},\
  \bibinfo {pages} {5124} (\bibinfo {year} {2018})}\BibitemShut {NoStop}%
\bibitem [{\citenamefont {Van~Wees}\ \emph {et~al.}(1992)\citenamefont
  {Van~Wees}, \citenamefont {de~Vries}, \citenamefont {Magn{\'e}e},\ and\
  \citenamefont {Klapwijk}}]{van1992excess}%
  \BibitemOpen
  \bibfield  {author} {\bibinfo {author} {\bibfnamefont {B.}~\bibnamefont
  {Van~Wees}}, \bibinfo {author} {\bibfnamefont {P.}~\bibnamefont {de~Vries}},
  \bibinfo {author} {\bibfnamefont {P.}~\bibnamefont {Magn{\'e}e}}, \ and\
  \bibinfo {author} {\bibfnamefont {T.}~\bibnamefont {Klapwijk}},\ }\href
  {\doibase https://doi.org/10.1103/PhysRevLett.69.510} {\bibfield  {journal}
  {\bibinfo  {journal} {Physical Review Letters}\ }\textbf {\bibinfo {volume}
  {69}},\ \bibinfo {pages} {510} (\bibinfo {year} {1992})}\BibitemShut
  {NoStop}%
\bibitem [{\citenamefont {Petrashov}\ \emph {et~al.}(1993)\citenamefont
  {Petrashov}, \citenamefont {Antonov}, \citenamefont {Delsing},\ and\
  \citenamefont {Claeson}}]{petrashov1993}%
  \BibitemOpen
  \bibfield  {author} {\bibinfo {author} {\bibfnamefont {V.~T.}\ \bibnamefont
  {Petrashov}}, \bibinfo {author} {\bibfnamefont {V.~N.}\ \bibnamefont
  {Antonov}}, \bibinfo {author} {\bibfnamefont {P.}~\bibnamefont {Delsing}}, \
  and\ \bibinfo {author} {\bibfnamefont {R.}~\bibnamefont {Claeson}},\ }\href
  {\doibase 10.1103/PhysRevLett.70.347} {\bibfield  {journal} {\bibinfo
  {journal} {Phys. Rev. Lett.}\ }\textbf {\bibinfo {volume} {70}},\ \bibinfo
  {pages} {347} (\bibinfo {year} {1993})}\BibitemShut {NoStop}%
\bibitem [{\citenamefont {G{\"u}l}\ \emph {et~al.}(2014)\citenamefont
  {G{\"u}l}, \citenamefont {Demarina}, \citenamefont {Bl{\"o}mers},
  \citenamefont {Rieger}, \citenamefont {L{\"u}th}, \citenamefont {Lepsa},
  \citenamefont {Gr{\"u}tzmacher},\ and\ \citenamefont
  {Sch{\"a}pers}}]{gul2014flux}%
  \BibitemOpen
  \bibfield  {author} {\bibinfo {author} {\bibfnamefont {{\"O}.}~\bibnamefont
  {G{\"u}l}}, \bibinfo {author} {\bibfnamefont {N.}~\bibnamefont {Demarina}},
  \bibinfo {author} {\bibfnamefont {C.}~\bibnamefont {Bl{\"o}mers}}, \bibinfo
  {author} {\bibfnamefont {T.}~\bibnamefont {Rieger}}, \bibinfo {author}
  {\bibfnamefont {H.}~\bibnamefont {L{\"u}th}}, \bibinfo {author}
  {\bibfnamefont {M.}~\bibnamefont {Lepsa}}, \bibinfo {author} {\bibfnamefont
  {D.}~\bibnamefont {Gr{\"u}tzmacher}}, \ and\ \bibinfo {author} {\bibfnamefont
  {T.}~\bibnamefont {Sch{\"a}pers}},\ }\href {\doibase
  https://doi.org/10.1021/nl502598s} {\bibfield  {journal} {\bibinfo  {journal}
  {Physical Review B}\ }\textbf {\bibinfo {volume} {89}},\ \bibinfo {pages}
  {045417} (\bibinfo {year} {2014})}\BibitemShut {NoStop}%
\bibitem [{\citenamefont {Albrecht}\ \emph {et~al.}(2017)\citenamefont
  {Albrecht}, \citenamefont {Moers},\ and\ \citenamefont
  {Hermanns}}]{albrecht2017hnf}%
  \BibitemOpen
  \bibfield  {author} {\bibinfo {author} {\bibfnamefont {W.}~\bibnamefont
  {Albrecht}}, \bibinfo {author} {\bibfnamefont {J.}~\bibnamefont {Moers}}, \
  and\ \bibinfo {author} {\bibfnamefont {B.}~\bibnamefont {Hermanns}},\ }\href
  {http://dx.doi.org/10.17815/jlsrf-3-158} {\bibfield  {journal} {\bibinfo
  {journal} {Journal of large-scale research facilities JLSRF}\ }\textbf
  {\bibinfo {volume} {3}},\ \bibinfo {pages} {112} (\bibinfo {year}
  {2017})}\BibitemShut {NoStop}%
\end{thebibliography}

\begin{thebibliography}{2}%
\makeatletter
\providecommand \@ifxundefined [1]{%
 \@ifx{#1\undefined}
}%
\providecommand \@ifnum [1]{%
 \ifnum #1\expandafter \@firstoftwo
 \else \expandafter \@secondoftwo
 \fi
}%
\providecommand \@ifx [1]{%
 \ifx #1\expandafter \@firstoftwo
 \else \expandafter \@secondoftwo
 \fi
}%
\providecommand \natexlab [1]{#1}%
\providecommand \enquote  [1]{``#1''}%
\providecommand \bibnamefont  [1]{#1}%
\providecommand \bibfnamefont [1]{#1}%
\providecommand \citenamefont [1]{#1}%
\providecommand \href@noop [0]{\@secondoftwo}%
\providecommand \href [0]{\begingroup \@sanitize@url \@href}%
\providecommand \@href[1]{\@@startlink{#1}\@@href}%
\providecommand \@@href[1]{\endgroup#1\@@endlink}%
\providecommand \@sanitize@url [0]{\catcode `\\12\catcode `\$12\catcode
  `\&12\catcode `\#12\catcode `\^12\catcode `\_12\catcode `\%12\relax}%
\providecommand \@@startlink[1]{}%
\providecommand \@@endlink[0]{}%
\providecommand \url  [0]{\begingroup\@sanitize@url \@url }%
\providecommand \@url [1]{\endgroup\@href {#1}{\urlprefix }}%
\providecommand \urlprefix  [0]{URL }%
\providecommand \Eprint [0]{\href }%
\providecommand \doibase [0]{https://doi.org/}%
\providecommand \selectlanguage [0]{\@gobble}%
\providecommand \bibinfo  [0]{\@secondoftwo}%
\providecommand \bibfield  [0]{\@secondoftwo}%
\providecommand \translation [1]{[#1]}%
\providecommand \BibitemOpen [0]{}%
\providecommand \bibitemStop [0]{}%
\providecommand \bibitemNoStop [0]{.\EOS\space}%
\providecommand \EOS [0]{\spacefactor3000\relax}%
\providecommand \BibitemShut  [1]{\csname bibitem#1\endcsname}%
\let\auto@bib@innerbib\@empty
\bibitem [{\citenamefont {Hikami}\ \emph {et~al.}(1980)\citenamefont {Hikami},
  \citenamefont {Larkin},\ and\ \citenamefont {Nagaoka}}]{hikami1980spin}%
  \BibitemOpen
  \bibfield  {author} {\bibinfo {author} {\bibfnamefont {S.}~\bibnamefont
  {Hikami}}, \bibinfo {author} {\bibfnamefont {A.~I.}\ \bibnamefont {Larkin}},\
  and\ \bibinfo {author} {\bibfnamefont {Y.}~\bibnamefont {Nagaoka}},\ }\href
  {https://doi.org/https://doi.org/10.1143/PTP.63.707} {\bibfield  {journal}
  {\bibinfo  {journal} {Progress of Theoretical Physics}\ }\textbf {\bibinfo
  {volume} {63}},\ \bibinfo {pages} {707} (\bibinfo {year} {1980})}\BibitemShut
  {NoStop}%
\bibitem [{\citenamefont {Lin}\ and\ \citenamefont {Bird}(2002)}]{Lin02}%
  \BibitemOpen
  \bibfield  {author} {\bibinfo {author} {\bibfnamefont {J.~J.}\ \bibnamefont
  {Lin}}\ and\ \bibinfo {author} {\bibfnamefont {J.~P.}\ \bibnamefont {Bird}},\
  }\href {http://stacks.iop.org/0953-8984/14/R501} {\bibfield  {journal}
  {\bibinfo  {journal} {Journal of Physics: Condensed Matter}\ }\textbf
  {\bibinfo {volume} {14}},\ \bibinfo {pages} {R501} (\bibinfo {year}
  {2002})}\BibitemShut {NoStop}%
\end{thebibliography}
\end{document}